\pdfoutput=1    
\documentclass[useAMS]{mn2e}
\usepackage{gensymb}
\usepackage{graphicx}  
\usepackage{mnras_cite}
\usepackage{ulem}
\usepackage{lineno}

\usepackage{dcolumn}  
\usepackage{bm}       
\usepackage{amssymb,amsmath}
\usepackage{color}

\usepackage{eso-pic} 

\AddToShipoutPicture*{
  \AtPageUpperLeft{
    \hspace{0.9\textwidth}
    \raisebox{-3.5\baselineskip}{
      \makebox[0pt][l]{\textnormal{DES-2015-0139}}
}}}

\AddToShipoutPicture*{
  \AtPageUpperLeft{
    \hspace{0.9\textwidth}
    \raisebox{-4.5\baselineskip}{
      \makebox[0pt][l]{\textnormal{FERMILAB-PUB-15-571}}
}}}
\newcommand{\bea}{\begin{eqnarray}}
\newcommand{\eea}{\end{eqnarray}}

\newcommand{\beq}{\begin{equation}}
\newcommand{\eeq}{\end{equation}}

\newcommand{\beqa}{\begin{eqnarray}}
\newcommand{\eeqa}{\end{eqnarray}}

\newcommand{\healpix}{{\tt Healpix}}

\def\nn{{\~n}}

\def\la{\mathrel{\mathpalette\fun <}}
 \def\ga{\mathrel{\mathpalette\fun >}}
\def\fun#1#2{\lower3.6pt\vbox{\baselineskip0pt\lineskip.9pt
\ialign{$\mathsurround=0pt#1\hfil##\hfil$\crcr#2\crcr\sim\crcr}}}

\begin{document}

\title[Angular clustering with photometric redshift PDFs]
{Galaxy clustering with photometric surveys using PDF redshift information}
\author[Asorey et al.]
{
\parbox{\textwidth}{J. Asorey$^{1}$\thanks{jasorey@illinois.edu}, M. Carrasco Kind$^{2,3}$, I. Sevilla-Noarbe$^{2,4}$, R. J. Brunner$^{1,2,3}$, J. Thaler$^{1,2}$
}	\vspace{0.4cm}\\
$^1$Department of Physics, University of Illinois, Urbana, IL 61801 USA\\
$^2$Department of Astronomy, University of Illinois, Urbana, IL 61801 USA\\
$^3$National Center for Supercomputing Applications, Urbana, IL 61801 USA\\
$^4$Centro de Investigaciones Energ\'eticas, Medioambientales y Tecnol\'ogicas (CIEMAT), Madrid, Spain}

\date{\today}
\pagerange{1--19} \pubyear{2016}
\maketitle


\begin{abstract}\\
Photometric surveys produce large-area maps of the galaxy distribution, but with less accurate redshift information than is obtained from spectroscopic methods.  Modern photometric redshift (photo-z) algorithms use galaxy magnitudes, or colors, that are obtained through multi-band imaging to produce a probability density function (PDF) for each galaxy in the map.  We used simulated data to study the effect of using different photo-z estimators to assign galaxies to redshift bins in order to compare their effects on angular clustering and galaxy bias measurements.  We found that if we use the entire PDF, rather than a single-point (mean or mode) estimate, the deviations are less biased, especially when using narrow redshift bins.  When the redshift bin widths are  $\Delta z=0.1$, the use of the entire PDF reduces the typical measurement bias from $5\%$, when using single point estimates, to $3\%$.

\end{abstract}


\section{Introduction}\label{sec:introduction}
The analysis of the three-dimensional distribution of galaxies has become one
of the major probes used to understand the history of the Universe and the growth of
matter perturbations. Spectroscopic surveys such as WiggleZ\footnote{http://wigglez.swin.edu.au/site/} \cite{wigglezover},
BOSS\footnote{https://www.sdss3.org/surveys/boss.php} \cite{bossover} and VVDS\footnote{http://cesam.oamp.fr/vvdsproject/vvds.html} \cite{vvds} 
have obtained precise maps of this distribution and many studies have increased our
knowledge about the expansion history of the Universe and the growth of
structures. However, targeting galaxies and obtaining spectra is a slow and costly process; therefore, past and current spectroscopic surveys 
have been limited to relatively low redshift and a reduced number of galaxies with respect to photometric surveys.

Multi-band imaging of wide areas of the sky is complementary to spectroscopic surveys.
These photometric surveys, such as the Canada-France-Hawaii Telescope Legacy Survey 
(CFHTLS)\footnote{http://www.cfht.hawaii.edu/Science/CFHTLS/},  
Dark Energy Survey (DES\footnote{http://www.darkenergysurvey.org/}, \pcite{desover}) and the Large 
Synoptic Survey Telescope (LSST\footnote{http://www.lsst.org/lsst/}, \pcite{lsstcol}), enable lower accuracy redshift estimation 
from the colours of millions of galaxies without being affected by spectroscopic selection effects. 

Photometric redshifts (photo-z) are estimated by using multi-band photometry as inputs to one or more different techniques that
map galaxy photometric properties into a redshift. These techniques can broadly be classified into two categories; the first is known as template-based
methods \cite{BPZ,lephare}, in which a set of calibrated galaxy spectral energy distributions (SEDs) is fit to the photometric data to find the one that best represents the observed fluxes.
The second category use a spectroscopic training set and machine learning 
algorithms, such as artificial neural networks \cite{collister04}, boosted decision trees \cite{gerdes10}, or prediction trees and random forests \cite{tpz}, to generate a photo-$z$ PDF estimate.

Probability density functions of various astronomical measurements have been used in cosmological analyses, 
for example, luminosity functions \cite{sheth07}, weak lensing \cite{mandelbaum2008}, cluster identification 
\cite{vanbreukelen09}, the real-space clustering of quasars \cite{myers09} and tomographic magnification \cite{Morrison2012}. 
However, a systematic analysis of the use of photometric redshift PDFs in galaxy clustering has not been 
performed, mostly due to the lack of reliable PDF estimation and its computational cost.

In this work, we study how the angular clustering of galaxies depends on the chosen photo-$z$ estimate and on the photo-$z$ bin width,
by using realistic simulations to compare clustering measurements based on photometric and spectroscopic redshifts. 
We also show how to include the full probability density when 
estimating the angular correlation functions, and we study the impact of using 
different photo-z PDF statistics (e.g., mean, mode, and median) on estimating the galaxy bias. 
In Section \ref{sec:methodology} we describe the methodology followed in the paper. We describe the angular clustering 
results and the fitting to galaxy bias in Section \ref{sec:results}, and we discuss these results in Section \ref{sec:discussion}. 
Finally, we summarize the main conclusions in Section \ref{sec:conclusion}.

\section{Methodology}\label{sec:methodology}

The standard approach to analyse galaxy clustering in photometric surveys begins with subdividing the 
catalogue into sub-samples selected by 'top-hat' photo-z redshift bins. 
The photo-z value used to determine if  a galaxy is in one or another bin is usually a point estimate of the photo-z PDF.
In this paper, we quantify how angular clustering analyses are affected by the 
choice of the specific photometric redshift estimator, including one that uses the full PDF information instead of 
photo-z point estimates. We address this by
measuring the clustering of a subsample of the DES-BCC Aardvark simulation mock galaxy 
catalogue \cite{busha2013}. We compared the clustering measurements given by different
photo-z estimators and when considering different redshift bin widths. As a specific test,
we quantify these differences by fitting theoretical galaxy correlation functions
in order to estimate the linear galaxy bias, \cite{kaiser84}, as a metric to evaluate which photo-z
estimator is more reliable \cite{coupon12,crocce15,soltan2015}.

\subsection{Simulation Data}
The mock galaxy catalogue considered here is the Aardvark v1.0 catalogue from the Blind 
Cosmology Challenge (BCC) simulations, developed for the DES.  
The catalogue is created from three $\Lambda$CDM N-body dark 
matter simulations, with sizes of $1050$ Mpc/h, $2600$ Mpc/h, $4000$ Mpc/h and 
$1400^3$, $2048^3$ and $2048^3$ particles, respectively. They were created using Gadget-2 \cite{gadget} and 
initial conditions given by CAMB \cite{cambt} and 2LPT \cite{2lpt}. The algorithm that populates the dark matter 
halos with galaxies, ADDGALS \cite{busha2013}, follows a prescription based on 
SubHalo Abundance Matching (SHAM) techniques, \cite{conroy2006,busha2013,reddick2013}. 
The final catalogue is complete 
down to $r<25$ and covers $1/4$ of the sky. Galaxy properties such as colour or luminosity are 
assigned by matching a spectroscopic training sample from the SDSS DR6 value added 
catalogue \cite{blanton05} at low redshift. This training is extrapolated to higher redshift 
matching the colour distribution to SDSS DR8 \cite{sdssdr8} and DEEP2 \cite{deep2} photometric data. Then,
the output catalogue includes DES colors and errors for each galaxy of the catalogue. 
These catalogues have been compared with real data by the DES collaboration 
\cite{chang2015,leistedt2015}.

The full BCC-Aardvark-v1.0c catalogue covers $10,313$ square degrees to the full DES depth, 
and includes a total of $1.36 \times 10^{9}$ galaxies. The simulated catalogue is stored in files 
according to \healpix \footnote{http://healpix.jpl.nasa.gov/} \cite{healpix} pixels of $n_{side}=8$. 
We chose a contiguous area of the simulation by using $24$ pixels, 
which corresponds to an area of about $1,200$ square degrees on the sphere in order to 
have a significant sampling of small scales.
 For our study, we have selected the galaxy sample according to a magnitude 
limited cut of $g<24$. This cut corresponds to a selection in the g-band of signal-to-noise greater than 20 
in the simulation, which incorporates the DES observed photometric error. 
The total number of galaxies in the catalogue after applying the 
magnitude cut is around 30 million galaxies. 

\subsection{Photometric redshift code}
We have used the publicly available code TPZ\footnote{http://lcdm.astro.illinois.edu/code/mlz.html} 
\cite{tpz} to estimate the galaxy 
redshift probability distributions. TPZ is a parallel code that estimates photo-$z$ PDFs using 
prediction trees and random forests. A prediction tree is constructed by splitting 
the data in recursive branches until a convergence criterion is reached. In order to 
construct more robust PDFs, the code uses the random forest technique in which $N_T$ 
bootstrap samples of the training set are created and prediction trees are generated for the 
$N_T$ samples. In order to include the measurement errors, e.g., magnitude errors, $N_R$ 
training samples are created by perturbing the training set according to the errors of the 
measurement variables. Finally, the PDF of each galaxy in the sample is created by 
combining the prediction trees. TPZ was one of the algorithms used in the 
DES Science Verification Data \cite{sanchez14}, and produced one of the best performances for that  dataset. 

We have considered $10^5$ galaxies as a training set, up to the full depth and a cut 
in the magnitude errors avoiding extremely large values in order to use all the available 
magnitude-redshift information from the simulation; and we, therefore, use less than $1\%$ 
of the available data for this purpose. The training set galaxies were confined to a region of $54$ square degrees. 
The test data used for the main analysis of this paper was directly selected from the simulation with no cuts on magnitude errors. 
The effect of the redshift selection of galaxies for the training set 
in the results is shown in section \ref{sec:trainsel}. 

As defined in \pcite{tpz}, the concentration of individual galaxy 
PDFs, $p(z)$, is output by TPZ as a PDF concentration parameter called zConf. 
This parameter is defined as the integrated probability between $z_{phz}\pm\sigma_{TPZ}(1+z_{phz})$, where $z_{phz}$ is the photometric redshift, 
and it measures the narrowness of the PDF. In this case, we 
selected $\sigma_{TPZ}=0.075$, which is similar to the $1-\sigma$ confidence interval of the PDFs. We can select different quality cuts by using this parameter, which is 
related to the BPZ ODDS parameter \cite{BPZ}. 

TPZ is a particular method of the MLZ framework. MLZ is code that computes photometric 
redshift PDFs using machine learning techniques. It 
incorporates a Bayesian combination of techniques that estimate photometric redshift PDFs, including both 
template based methods and unsupervised machine learning algorithms \cite{bayesian}, and 
also enables an efficient storage of the PDFs by using a sparse representation basis \cite{sparse}.  
For simplicity, we only used TPZ for the photometric redshifts in this paper, which is justified by the excellent results produced by 
TPZ on the DES Science Verification Data \cite{sanchez14}, and by the fact that we want to study the impact of using 
photo-$z$ PDFs in clustering as produced by a single technique (to simplify the resulting analysis). TPZ has been used, together with other codes listed in \pcite{sanchez14}, in 
several DES Science Verification Data studies \cite{bonnet15,crocce15,giannantonio15,des15}.

\subsection{Survey configuration: Photo-z binning}
\begin{figure}
\begin{center}
\includegraphics[trim = 0cm 0cm 0cm 0cm, width=0.43\textwidth]{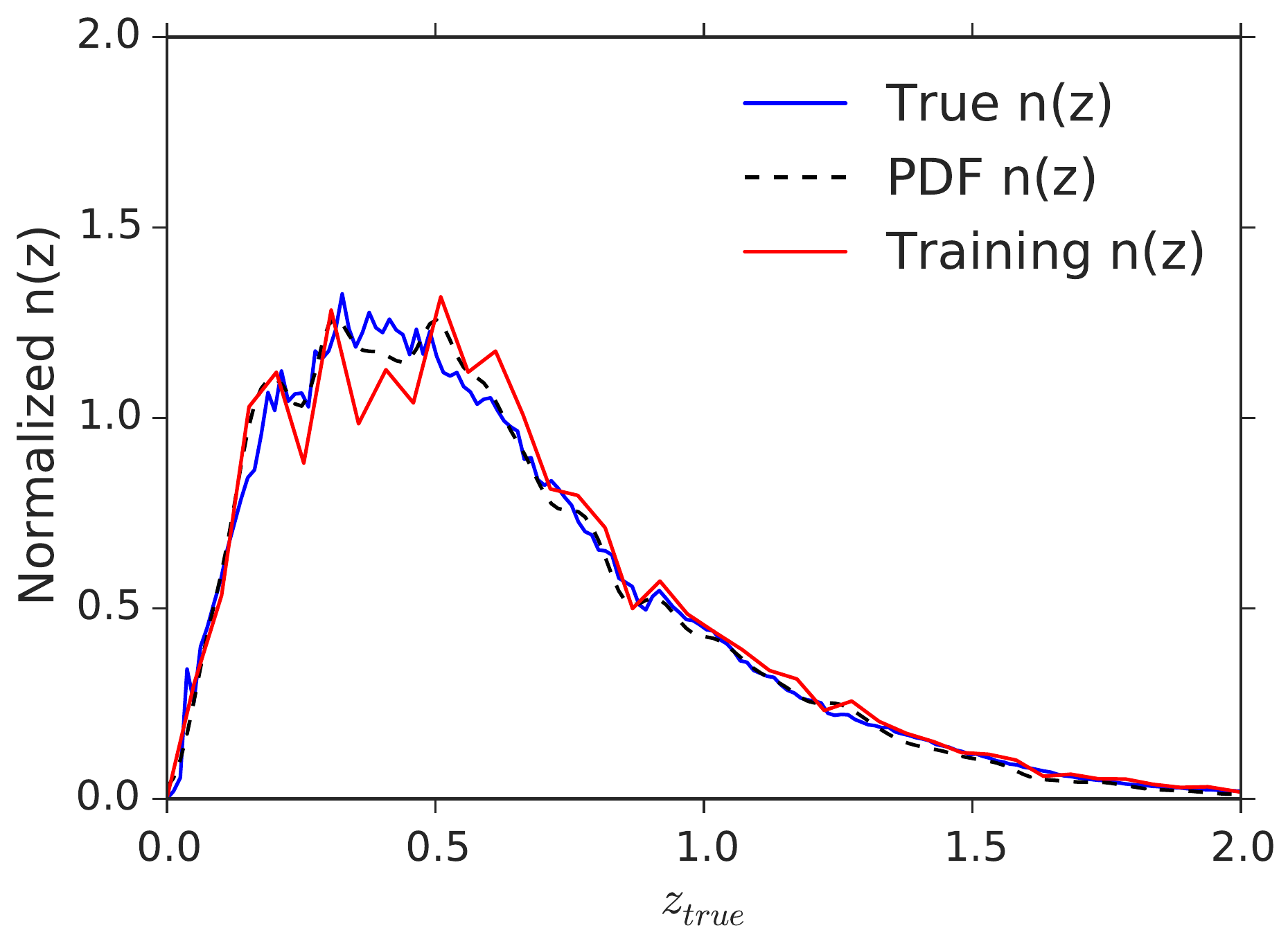}
\caption{In blue, the galaxy redshift distribution for 31 million galaxies with $g<24$ in the BCC Aadvark 1.0 catalogue
 in the selected region of $1,200$ square degrees used for our analysis. The dashed black line shows the 
 result of stacking the individual PDFs of all the galaxies in the sample. In red, we show the true $n(z)$ for the a subset of galaxies with $g<24$ of the training set used in 
 the photo-z machine learning algorithm.}
\label{fig:fig2}
\end{center}
\end{figure}

Galaxy clustering analyses in photometric surveys are usually done by measuring angular correlations of 
galaxy samples selected in different redshift bins. We divide the full redshift range, which in this paper we restrict to the range $0.2<z<1.4$, into 
$N_z$ redshift bins of width $\Delta z$ in order to reduce the extent of the projection of radial information for 
2D clustering analysis. As shown in \pcite{Asorey2012} and \pcite{eriksen15}, the optimal photometric redshift bins 
are given by shells of about twice the size of the photometric redshift standard deviation. In this paper, we consider different configurations: $\Delta z=\{0.1,0.15,0.2,0.3\}$ in order to study the evolution of photometric clustering 
with bin width. The true redshift distribution of galaxies, $n(z)$, is shown in Figure \ref{fig:fig2}, together with the redshift distribution 
obtained by stacking TPZ PDFs. We also show (red) the $n(z)$ of the spectroscopic training set.
For this paper, we were only interested in a patch of the sky that covers $1,200$ square 
degrees, which allows us to measure the small scale clustering and to study how it depends on 
the different photo-z statistical quantities. 

\begin{figure}
\begin{center}
\includegraphics[trim = 0cm 0cm 0cm 0cm, width=0.43\textwidth]{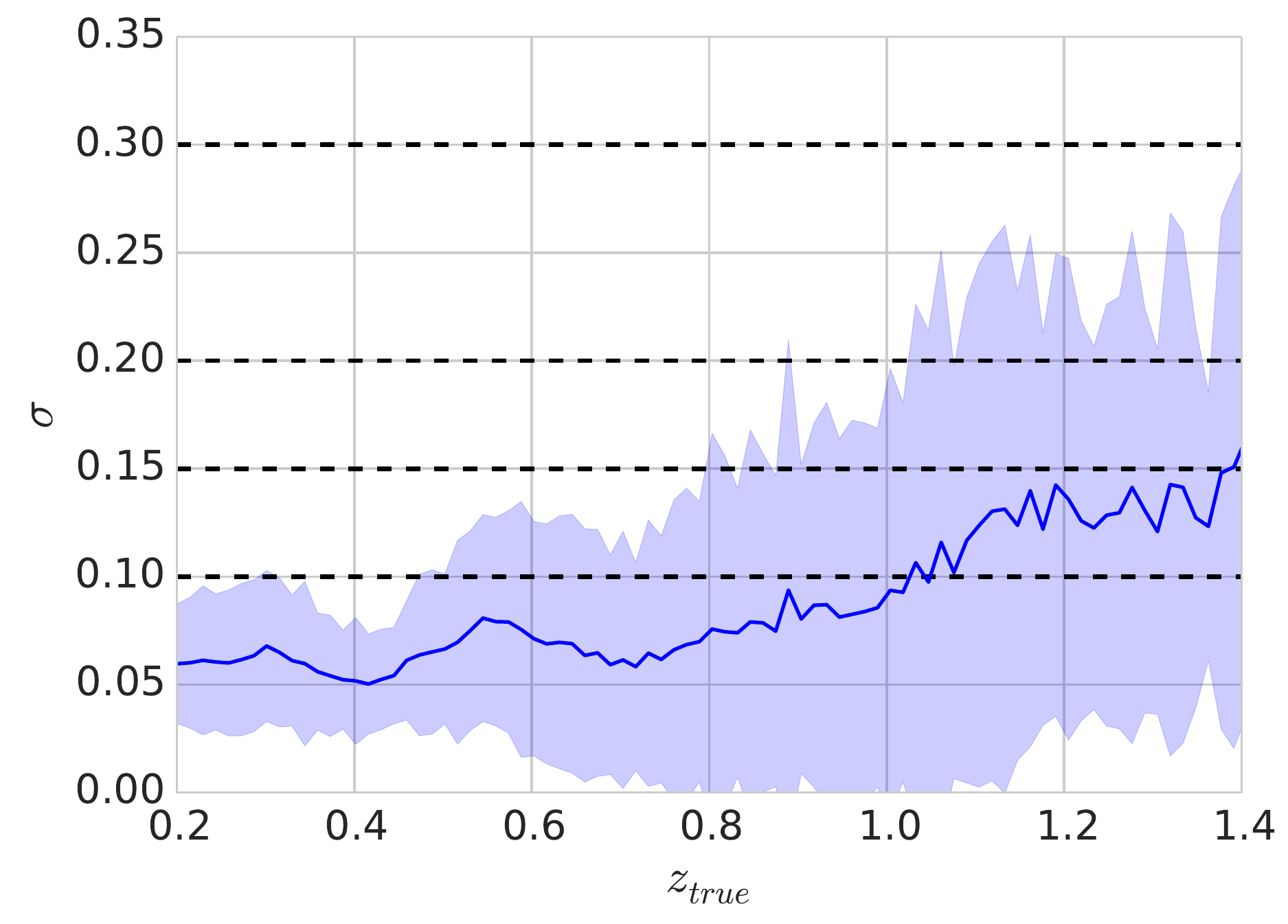}
\caption{Dependence of the square root of the mean photo-z variance, given by $\sigma^2$ in Equation (\ref{eq:sigmaphz}) for galaxies in small true redshift bins, on the true redshift. Standard deviations are given by the square root of the variance of the photo-z errors in each bin. Dashed lines correspond to the width of the different bin configurations treated in this paper, in order to compare the bin widths with the photo-z dispersion for the sample considered.}
\label{fig:fig3}
\end{center}
\end{figure}

\begin{figure}
\begin{center}
\includegraphics[trim = 0cm 0cm 0cm 0cm, width=0.43\textwidth]{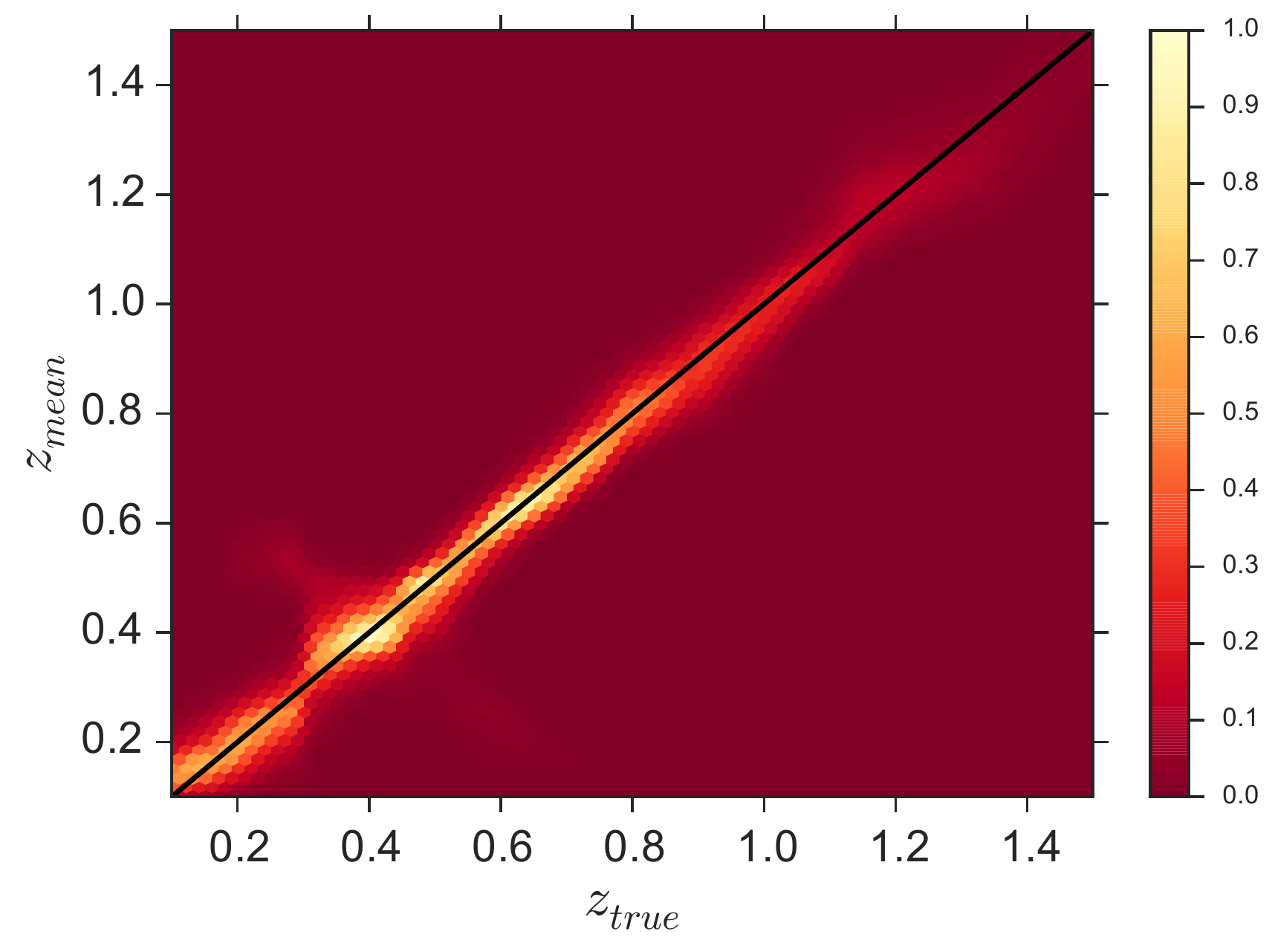} 
\caption{Relative number of galaxies with mean photometric redshift $z_{mean}$ and true redshift $z_{true}$. It contains the information of the dispersion of mean photometric redshifts with respect to the true redshift. The color code corresponds to the relative number of galaxies with 
respect to the 1:1 relation (black line) between true redshift and mean photometric redshift.}
\label{fig:fig3b}
\end{center}
\end{figure}

In Figure \ref{fig:fig3}, we present the evolution with redshift of the photometric redshift error, $\sigma$, for the BCC sample of galaxies, 
with $g<24$, given by: 
\begin{equation}
\sigma^2=\int{(z-\bar{z})^2p(z)dz}
\label{eq:sigmaphz}
\end{equation} 
where $\bar{z}$ is the mean redshift, defined in Equation (\ref{eq:mean}) below, and we compare it with the different redshift bin widths that we have considered in this paper. Although the optimal choice would be $\Delta z=0.15$ or $0.2$, we also consider extreme cases with $\Delta z=0.1$ or $0.3$ to extend the analysis of the dependence of photo-z clustering on this quantity. In Figure \ref{fig:fig3b}, we show the normalized dispersion between the true redshift and the mean redshift. 

\subsection{Photo-z estimators per galaxy}\label{sec:estimators}
\subsubsection{Single point statistics}
Once we have computed photo-z PDFs with TPZ, we estimate single statistical summary quantities. In this study, we focus on the mode redshift, $\hat{z}$, and the mean redshift, $\bar{z}$. 

We define the mean as the first moment of the PDF, $p(z)$:
\begin{equation}
\bar{z}=\int{zp(z) dz}
\label{eq:mean}
\end{equation}
The mode redshift is the redshift with highest probability in the PDF, $p(z)$:
\begin{equation}
p(\hat{z})=p_{max}
\label{eq:mode}
\end{equation}
As the output of the PDF is binned in 200 bins the used "mode" corresponds to the redshift of the bin with the highest probability.
Another summarization or single point estimate that we consider in the paper is the Monte Carlo sampling redshift, $z_{MC}$ \cite{wittman2009}. 
The Monte Carlo photo-z is the redshift that corresponds to the value of the cumulative distribution function given by a random number in the interval $(0,1]$. 

We also evaluated the median redshift, but we decided not to include it in the final results as it is similar to the other point estimates, thus for clarity we decided to 
reduce the analysis to the chosen single point estimates. We created a catalogue for each redshift bin considered by selecting the galaxies with the single point estimate in the range covered by the redshift bin.

\subsubsection{Photo-z weights}
The proposed technique to incorporate the PDF information in our analysis consists on doing number 
counts in redshift bins according to a weight 
for each galaxy in each radial shell, where the weight is given by the probability that the galaxy lies in the 
corresponding redshift bin. 
Because the TPZ PDF output is discretized the PDF is given in redshift bins. The output is normalized such that   
$\sum{p_k}=1$, where $p_k$ is the probability for the k-bin. 
We define the galaxy weight in a redshift bin $z_{min}<z<z_{max}$ as:
\begin{equation}
f_z = \sum_{k}{p_k(z_k)}
\label{eq:weight_d}
\end{equation}
where we add the values for redshifts $z_k \in [z_{min},z_{max}]$ that belong to the redshift bin in consideration. 
According to this definition, a galaxy may have weights in different redshift bins, where the
total weight of the galaxy in the whole redshift space is $f_{tot}=\sum_{j=1}^{N_z}{f_{z_j}}=1$. We measured the 
galaxy clustering by using the weights for the galaxy counts. The case that involves the photo-z single 
point estimates (mean, mode) is equivalent to setting the weight to  $f_z=1$ for all galaxies selected 
in the corresponding redshift bin.

We defined threshold cuts, $p_{threshold}$, in a similar way as in \pcite{mandelbaum2008}, 
as the process of determining if a galaxy lies within a redshift bin or not when using weights. Thus, a galaxy 
$\alpha$ in redshift bin $j$ would only be incorporated if
\begin{equation}
f_z > p_{threshold}
\label{eq:threscut}
\end{equation}
When PDFs are broad and contain multiple peaks, we might be introducing noise in each redshift bin from galaxies that are not in the bin but have 
a non-negligible weight. This can be addressed by applying the threshold cut.

\begin{figure}
\begin{center}
\includegraphics[trim = 0cm 0cm 0cm 0cm, width=0.43\textwidth]{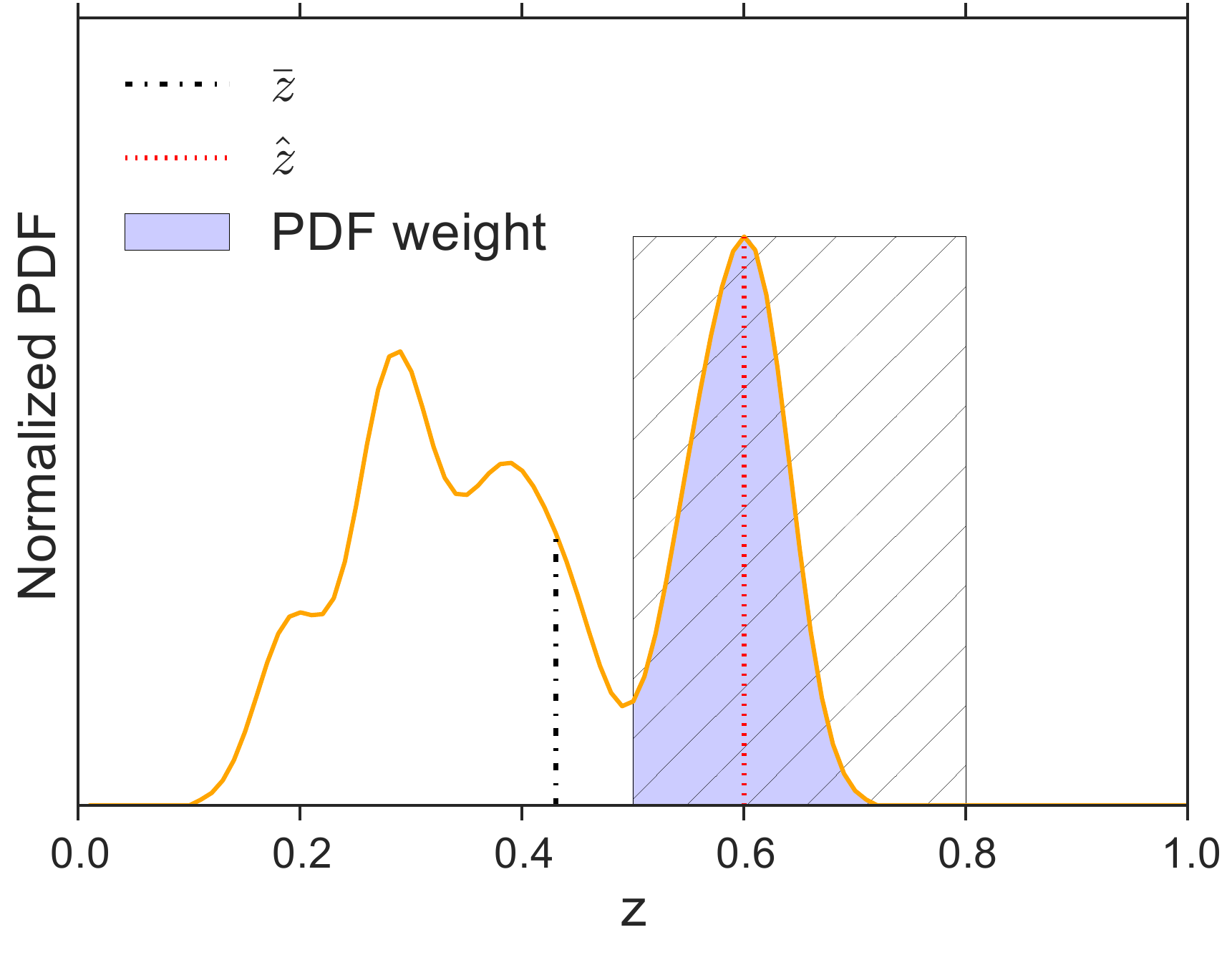}
\caption{Example of the definitions of the different photometric 
redshift estimators, where the mode redshift is shown by the vertical red dotted line and the mean redshift by the black dashed line.  The blue region corresponds to the part of the PDF that is between the photometric bin 
$0.5<z<0.8$ (true redshift is z=0.603), which is shown as a hatched area. The PDF weight defined in eq. (\ref{eq:weight_d}) would be the fraction of the total area below the continuous line that is contained in the blue region.}
\label{fig:fig1}
\end{center}
\end{figure}

In Figure \ref{fig:fig1}, we present a graphic example of the different photometric redshift estimators that we use. 
We intentionally selected a PDF with a most frequent (mode) redshift, given by eq. (\ref{eq:mode}) within the photometric bin $0.5<z<0.8$ but where the mean, defined in eq. (\ref{eq:mean}) is in a different redshift bin. In blue, we show the portion of the PDF between $0.5<z<0.8$ that corresponds to the weight of the galaxy in that redshift bin.
Of course the PDF displayed in Figure \ref{fig:fig1} is an extreme case. For this particular PDF, zConf=0.37, while near z=0.6 the mean PDF quality parameter is zConf$\sim0.95$.
In Appendix \ref{sec:ap1}, we discuss the statistical properties of the mean and the mode and the overall quality of the photometric sample PDFs . 

With photo-$z$ PDFs we can easily obtain the photometric sample redshift distribution, $n(z)$,  in each bin by stacking all the individual $p(z)$ of the selected galaxies.
\begin{equation}
n(z)_{pdf}=\sum{p_i(z)f_{z_i}}
\end{equation}
In this paper, we have considered this definition as the default estimation of $n(z)$ by setting $f_z=1$ for the galaxies selected according to single point statistics 
in a redshift bin. We can also determine the true $n(z)$ measured by the distribution of the true redshifts of this simulated sample. We weighted the true redshift of each galaxy by the PDF weight when considering full PDF information. Throughout this paper, when we refer to true values, we are considering the latter definition.


\subsection{Two point angular correlation function estimators}
\subsubsection{Pixel based estimator}
We computed the angular correlations by using pixel maps of the galaxy density field. These maps are created by using {\tt Healpix} for each redshift 
bin and for each photometric redshift estimator, with  $n_{side}=1024$, corresponding to a minimum angular resolution of 0.06 degrees. For the definition of the estimator, see 
\pcite{scranton02}, \pcite{crocce11}, and \pcite{wang13}. The angular correlation, is:
\begin{equation}
w(\theta) = \frac{1}{N_{pairs}}\sum_{ij}{\delta_i \delta_j}
\label{eq:wobs}
\end{equation}
where $N_{pairs}$ is the number of pixel pairs at an angle $\theta$. We defined the density contrast as 
$\delta_i=(n_i-\bar{n})/\bar{n}$ where $n_i$ is the number of counts in pixel $i$ and $\bar{n}$ the mean 
number density of galaxies. When selecting galaxies in terms of the PDFs, the total number of counts in every pixel is the sum of 
the weights of all the galaxies in the pixel, i.e.: $n_i=\sum\limits_{gal \in i}{f_z}$ and $\bar{n}=\sum\limits_i\sum\limits_{gal \in i}{f_z}$.

In our analysis, we only focused on individual redshift bins and we have not considered the correlations and covariance between different bins and the effect that 
assigning weights of the same galaxy to different bins might have in the analysis of galaxy clustering cross-correlations.
In order to include errors on our measurements, we considered jackknife samples, dividing the survey area into $N_{JK}$ regions, each about $3$ square degrees. The covariance matrix, therefore, is given by:
\begin{equation}
C_{\theta_i,\theta_j} = \frac{N_{JK}-1}{N_{JK}}\sum_{k=1}^{N_{JK}}(w_k(\theta_i)-w(\theta_i))(w_k(\theta_j)-w(\theta_j))
\end{equation}
which is the same definition used in \pcite{scranton02,norberg09,wang13}. For the galaxy bias fitting, we adopt the mixed approach used in \pcite{crocce15}, where the correlation matrix between diagonal elements and off-diagonal elements of the covariance matrix  is calculated by using theoretical angular power spectra that are rescaled by the variances given by the jackknife errors in order to determine the covariance matrix of the angular correlations.	

\subsubsection{Direct pair counting estimator}
An alternative method to measure angular correlations consists of using pair counts. In order to estimate the angular correlation functions, we used the Landy-Szalay estimator, \cite{landyszalay},
\beq
w(\theta)=\frac{DD-2DR+RR}{RR}
\label{eq:wtheta_p2p}
\eeq
where $DD$ is the number of galaxy-galaxy pairs, $DR$ is the galaxy-random pairs and $RR$ the random random pairs within $\theta$ and $\theta + \delta\theta$. Random catalogues are created by throwing points in the survey footprint following a uniform density. These are appropriately normalized to the total number of counts. When counting pairs, each galaxy was weighted according to Equation (\ref{eq:weight_d}).  We computed the point-to-point angular correlation functions by using the publicly 
available tree code\footnote{http://lcdm.astro.illinois.edu/code/tpacf.html},
explained and used in \pcite{dolence08,wang13}. We compare in section \ref{sec:results} the point-to-point estimator with the pixel based estimator.

\subsection{Theoretical modeling}
The angular auto-correlation within a given redshift bin is given by:
\begin{equation}
w(\theta)=\int{dr_1\phi(r_1)}\int{dr_2\phi(r_2)\xi(r_1,r_2,\theta)}
\label{eq:wtheta}
\end{equation}
where the spatial correlation function $\xi(r_1,r_2,\theta)$ encodes the 3D information of the density
field that we are projecting. The window functions,  $\phi$, are
a combination of the galaxy redshift distribution, $n(z)$, the galaxy bias, $b(z)$, and the linear growth rate of structure, 
$D(z)$, in such a way that $\phi(z(r))=n(z¨)b(z¨)D(z¨)$, where we assumed the linear local bias model \cite{kaiser84}:
\begin{equation}
\delta_g=b_g\delta 
\label{eq:bias}
\end{equation}
We parametrize the bias by one parameter $b$ per redshift bin in the following way:
\begin{equation}
w(\theta)=b^2\int{dz_1n(z_1)D(z_1)}\int{dz_2n(z_2)D(z_2)\xi(r_{12},\theta})
\label{eq:wthetadetail}
\end{equation}
where $r_{12}^2=r(z_1)^2+r(z_2)^2-2r(z_1)r(z_2)cos(\theta)$, being $r(z)$ the comoving distance 
to redshift $z$.

We used CAMB \cite{cambt} to obtain the linear power spectrum with 
Halofit \cite{halofit,takahashi2012} in order to include non-linearities at small scales. We Fourier transform this angular power spectrum in order to compute the 
3D angular correlations required by Equation (\ref{eq:wtheta}). For this paper, we considered a flat $\Lambda$CDM model driven by the simulation 
cosmological parameters when computing the theoretical correlation function.
We also included linear redshift space distortions as a series of multipoles following \cite{kaiser87,hamilton92}.

\begin{figure}
\begin{center}
\includegraphics[trim = 0cm 0cm 0cm 0cm, width=0.45\textwidth]{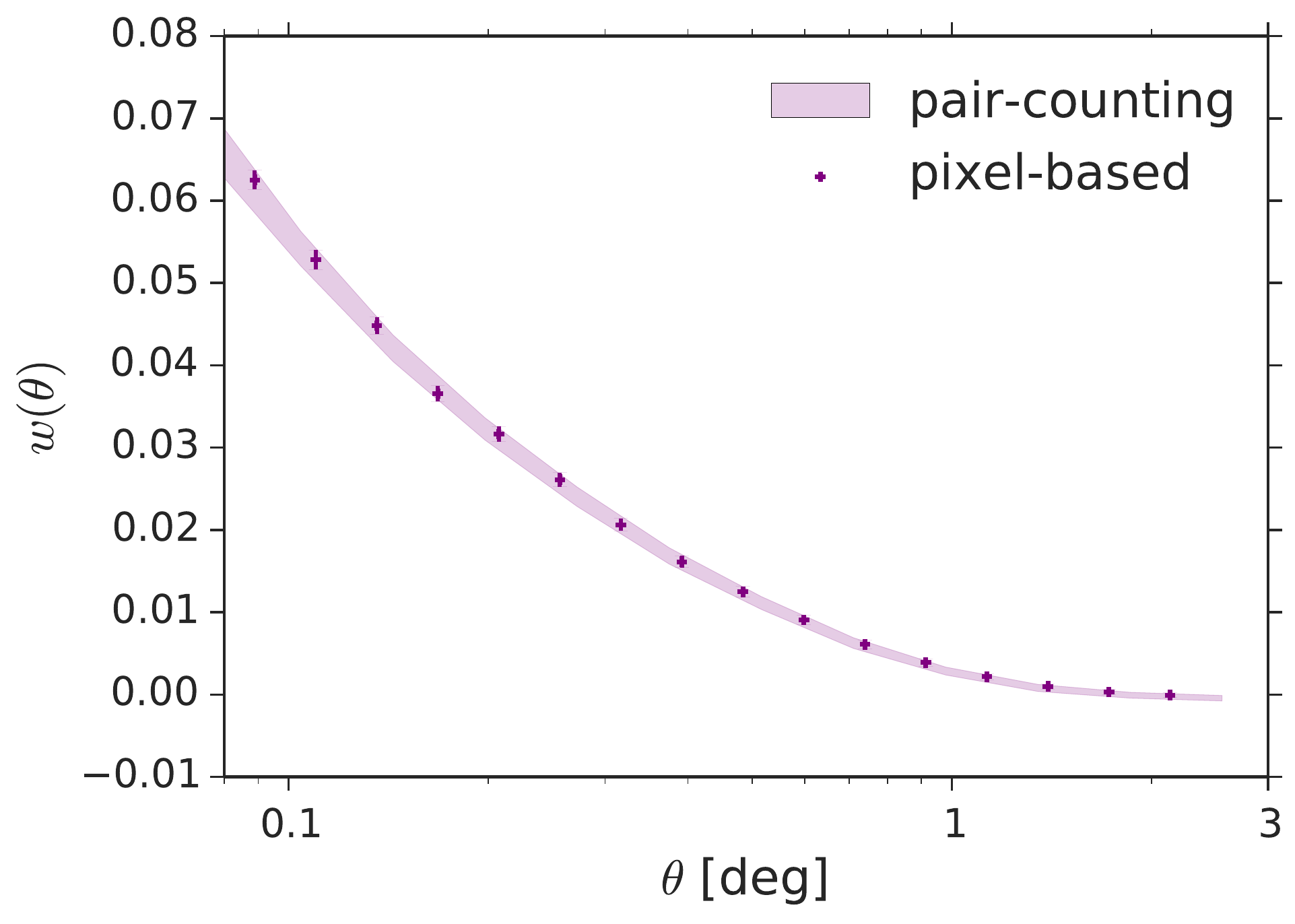}
\caption{A comparison, demonstrating good agreement, between the pixel based clustering measurement (purple points) and the 
point-to-point clustering measurement (purple shadow include measurements within the error bars) for the photometric redshift bin $1.0<z<1.2$ when considering galaxy weights and 
a threshold on the weights of $f_z>0.1$.}
\label{fig:fig4}
\end{center}	
\end{figure}

\begin{figure*}
\begin{center}
\begin{tabular}{cc}
\includegraphics[trim =0cm 0cm 0cm 0cm, width=0.8\textwidth]{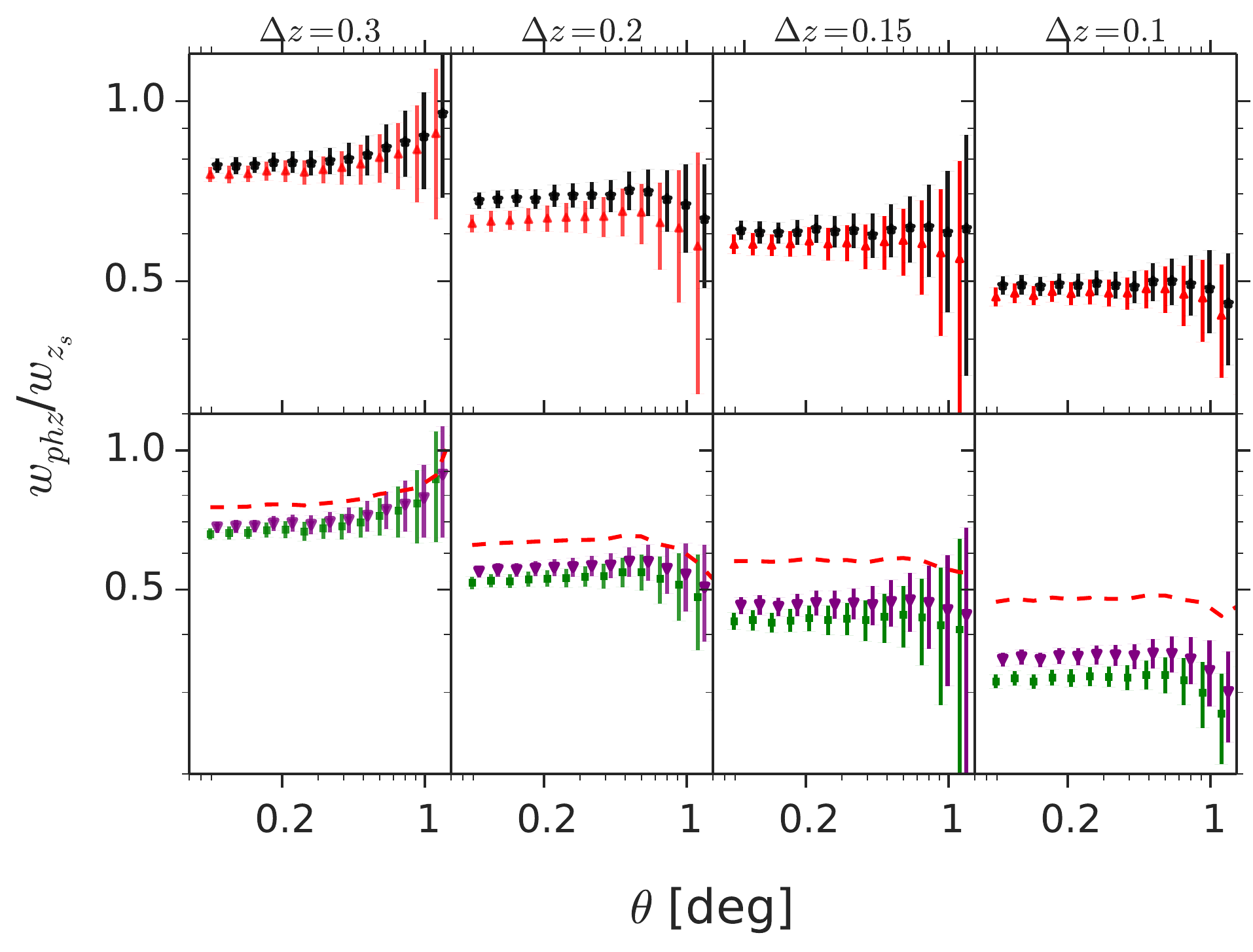}
\end{tabular}
\caption{This comparison between angular clustering measurements that use photometric redshifts and true redshifts. We consider different redshift estimators for redshift bins centered at $z=1$ but with different widths. (Top) The results when computed by using the mode (red triangles) and the mean (black stars). (Bottom) The results when using PDF weights, with threshold $p_{threshold}=0$  (green squares), and when considering thresholds to the weights of $p_{threshold}=0.1$ (purple triangles). In both panels, the different columns correspond to different bin widths. In the bottom row, we show the results for $\hat{z}$ with a red dashed line in order to compare with the upper panels. The ratios with respect to the true redshift results decrease with the bin width and are different depending on the considered photo-z statistics used. We have  shifted the x-axis positions of the black stars and purple triangles for clarity.}
\label{fig:fig5}
\end{center}
\end{figure*}

\section{Results}\label{sec:results}

\subsection{Comparison between direct pair counting and pixel-based estimators}\label{sec:res1}
As shown in \pcite{wang13}, pixel-based and point-to-point pair count methods yield similar results for 2-point angular clustering. However, this previous work only considered unweighted pair counts in any given bin, i.e., weights of $f_z=1$. In Figure \ref{fig:fig4}, we extend this previous result to compare the results for both pair count and pixel-based methods when considering galaxy 
weights in the redshift bin $1.0<z<1.2$ and a threshold $p_{threshold}=0.1$. As shown in this figure, over the angular range $0.1<\theta<1.0$, we find a good agreement. The number of jackknife regions is 
different in the two cases, however, being $N_{JK}=32$ for the point-to-point case and $N_{JK}=384$ when considering the pixel-based estimator. We opted to use the computationally simpler pixel-based estimator in the rest of the paper.

\subsection{Clustering amplitude}\label{sec:res2}
We now focus our analysis on the relative amplitude between the angular clustering signal using different photo-z selection criteria and the 
true redshift clustering. This allows us to directly study how the different statistical representations of photometric redshift change the measurement signal.

In the analysis, we divide the redshift range $0.2<z<1.4$ into different numbers of bins in order to consider different bin configurations. As a result, a comparison between individual redshift bins for different bin configurations may consider different redshift regions for different configuration. For this reason, we present clustering results for bins with different widths in symmetrical manner about a given central redshift. As discussed previously, we restrict our analysis to four different bin widths. In Figure \ref{fig:fig5}, we present the ratios of the photometric redshift clustering with respect to the true redshift clustering, for different statistical estimators and a redshift bin centered at $z=1$.

In the left panels, we show the results for a broad bin of $\Delta z=0.3$ in the redshift range $0.85<z<1.15$. As expected, for all algorithms, the amplitude of the clustering of the photometric sample is smaller than when using the true redshifts, since the errors on a photometric redshift estimate will suppress this inherent clustering.  Notice that the clustering measurements when using the mode, $\hat{z}$, and mean, $\bar{z}$,
 are similar. 
Any small differences are due to the fact that each selection produces a different $n(z)$ when individual PDFs are not symmetric, like the one shown in Figure \ref{fig:fig1}.  

In the bottom panels we show the clustering ratios with respect to the true clustering when using PDF weights with different probability thresholds. This allows us to both clean our sample, as if using a quality cut, and sample  
more narrow redshift bins. The thresholds considered are $p_{threshold}={0.0,0.1}$. The signal depends on the cut on the selection of 
weighted galaxies. The stronger the cut, the cleaner the sample and the clustering amplitude increases, as well as the intrinsic $n(z)$. But this may also bias our results as we may be changing the average galaxy types in the sample, as discussed in \pcite{marti14}. Here we considered a low threshold that is non-negligible to compare with the full PDF case. To aid in the comparison with the point estimate photo-z selection, we also present the ratio for $\hat{z}$ with the red dashed line. 

We consider narrower bin configurations in order to test what happens as we approach the intrinsic photo-z dispersion error, summarized in Figure \ref{fig:fig3}. 
We see in the case when $\Delta z=0.2$ that the amplitude of the photometric samples clustering decreases with respect to the  
true redshift clustering. The angular clustering signal is proportional to $n(z)^2$, as explained by Equation (\ref{eq:wtheta}), which in the top hat case means that it is inversely proportional to $(\Delta z)^2$. Photometric samples distribution in true redshift are broader than the top hat bin and therefore, the signal amplitude is smaller. The bin considered in this case for 
galaxy selection is  $0.9<z<1.1$. We see that the differences between the mode and the mean are bigger than in the previous case with $\Delta z = 0.3$. This may be a result of the fact that when we consider bins much bigger than the intrinsic separations, the differences between photo-z single statistic estimators are smaller.
As we extend the comparison to smaller widths, $\Delta z=0.15$ ($0.925<z<1.075$), the ratio between the angular clustering signal for the photometric samples and the sample with true redshift becomes smaller. This is in agreement with the trend we saw before,  as $w(\theta)\propto (1/\Delta z)^2$ for true redshift clustering, while the photo-z dispersion keeps the corresponding signal diluted in the radial direction.

The case with $\Delta z=0.1$ ($0.95<z<1.05$) shows the same trend than the previous cases with bigger bin widths. The ratio of the photometric redshift signals to the true redshift signal continues to decrease. The results with mean and mode estimators tend to converge as the bin width approaches the photo-z dispersion error. 

\begin{figure}
\begin{center}
\includegraphics[trim = 0cm 0cm 0cm 0cm, width=0.45\textwidth]{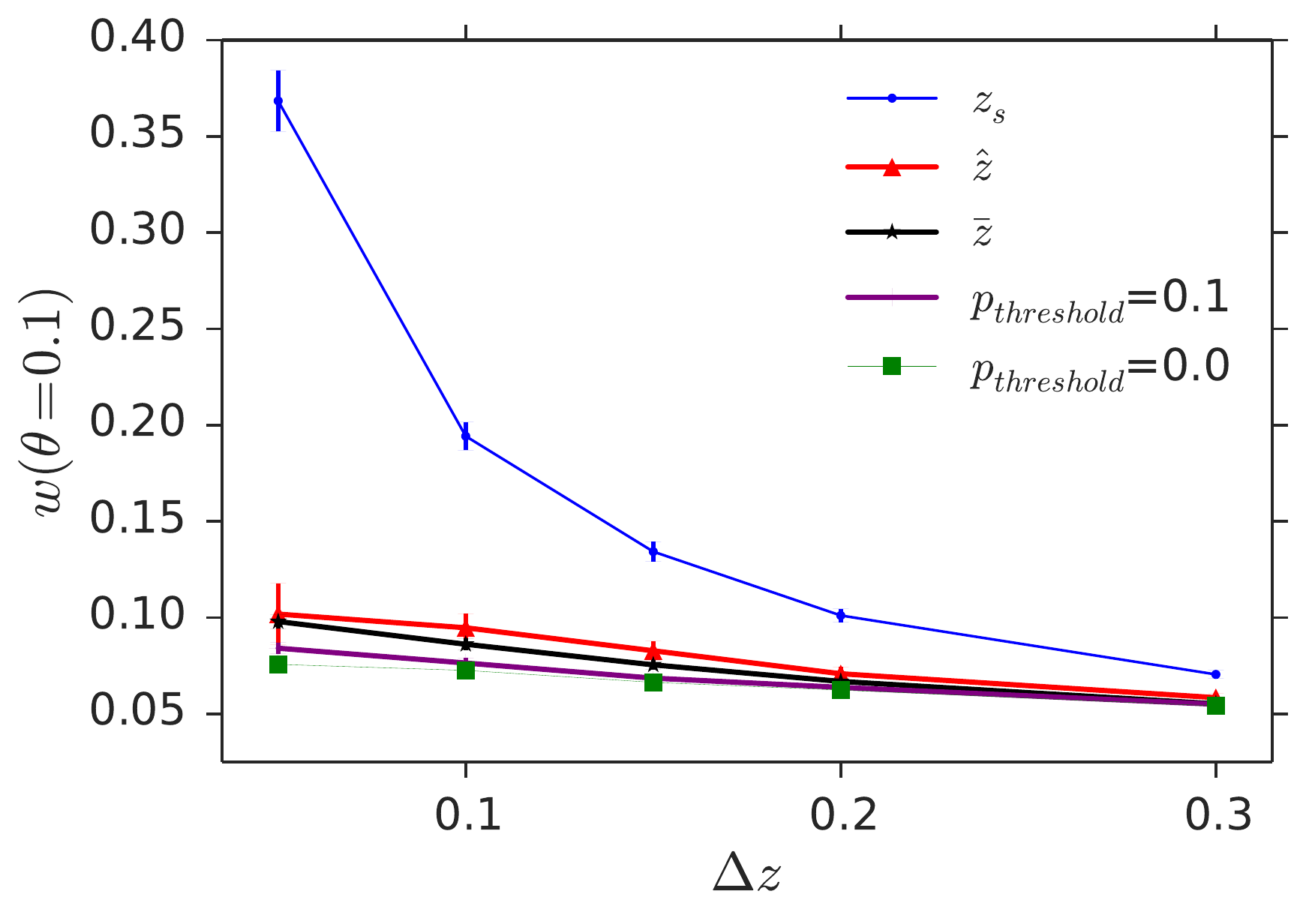}
\caption{The evolution of the clustering amplitude at $\theta=0.1$ for a redshift bin centered at $z=0.5$ with a given bin width, $\Delta z$. We show the evolution for the true redshift 
 and for different photometric redshift statistics in order to quantify when the clustering signal saturates with bin width.} 
\label{fig:fig5b}
\end{center}	
\end{figure}
\begin{figure*}
\begin{center}
\begin{tabular}{cc}
\includegraphics[trim = 1cm 0cm 0cm 0.5cm, width=0.48\textwidth]{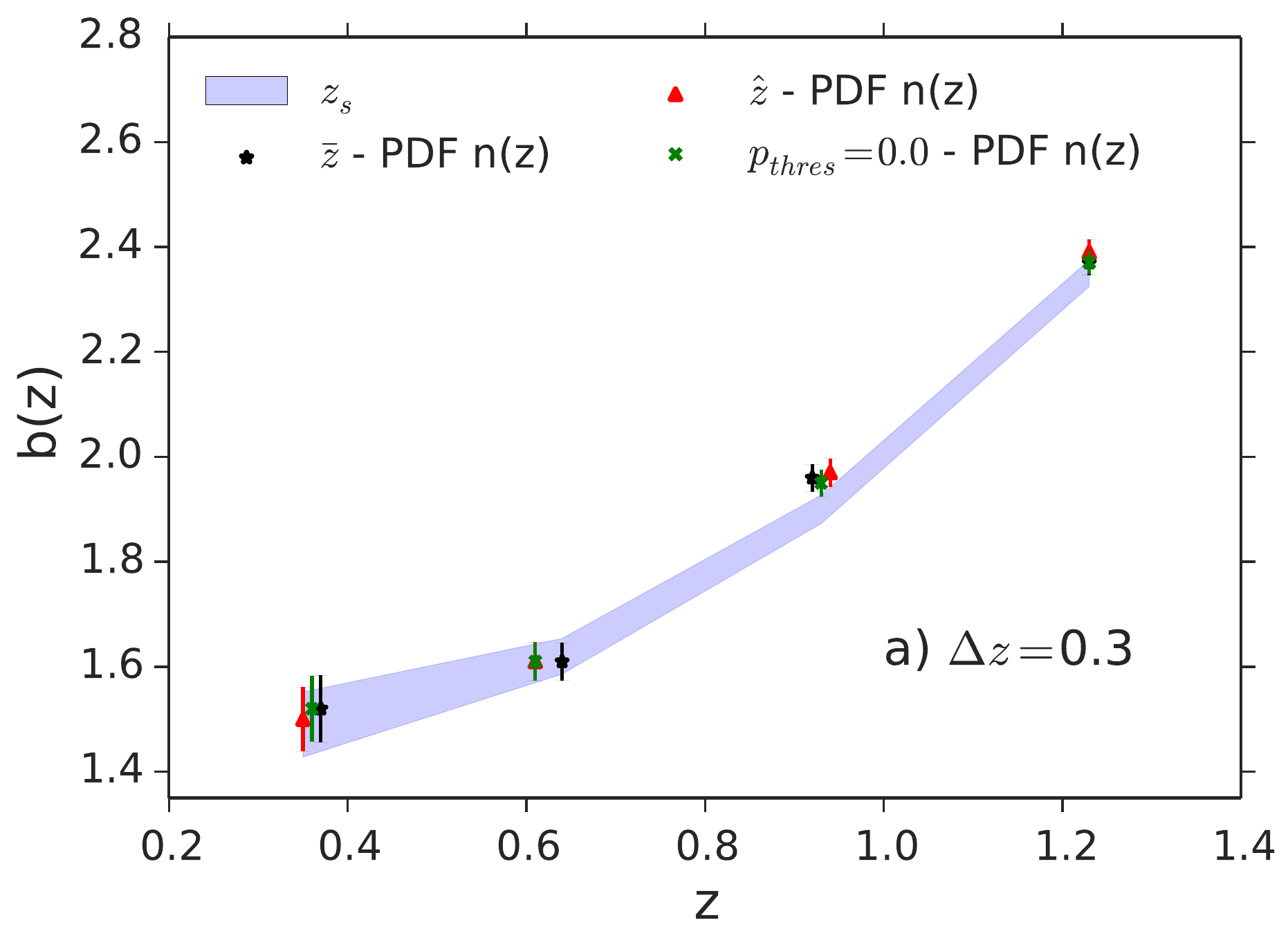} 
\includegraphics[trim = 0cm 0cm 1cm 0.5cm, width=0.48\textwidth]{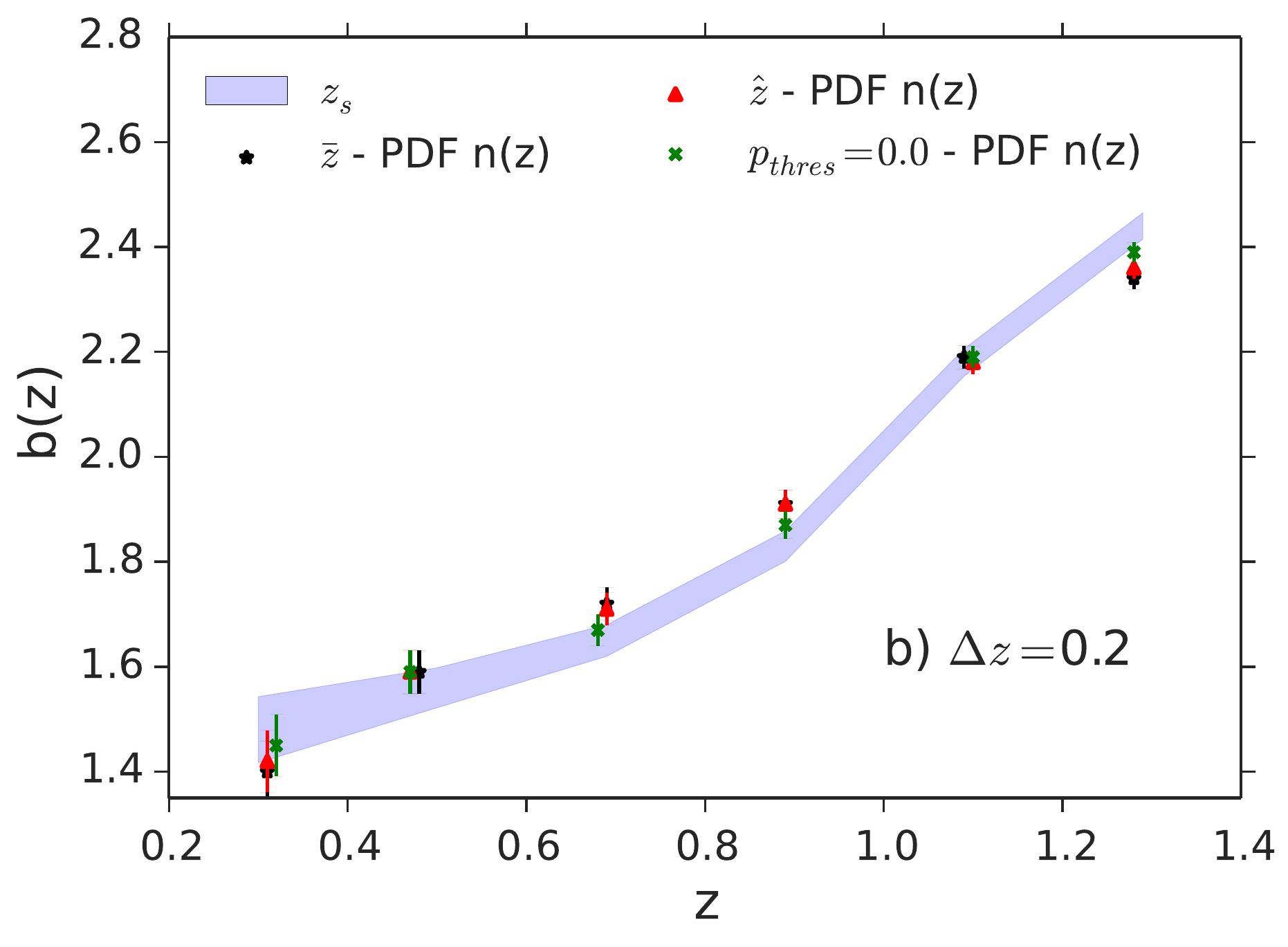} \\
\includegraphics[trim = 1cm 0cm 0cm 0.5cm, width=0.48\textwidth]{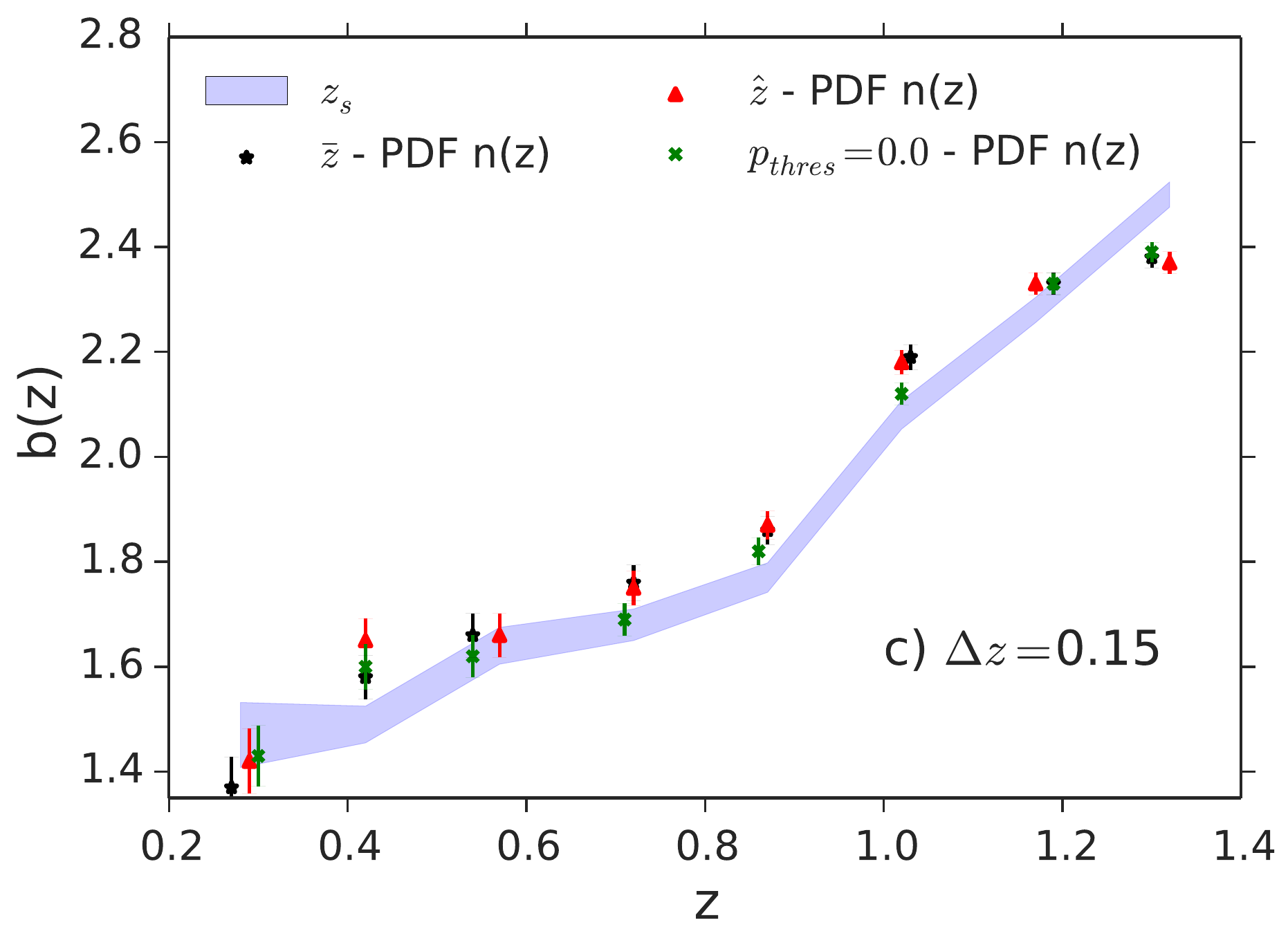}
\includegraphics[trim = 0cm 0cm 1cm 0.5cm, width=0.48\textwidth]{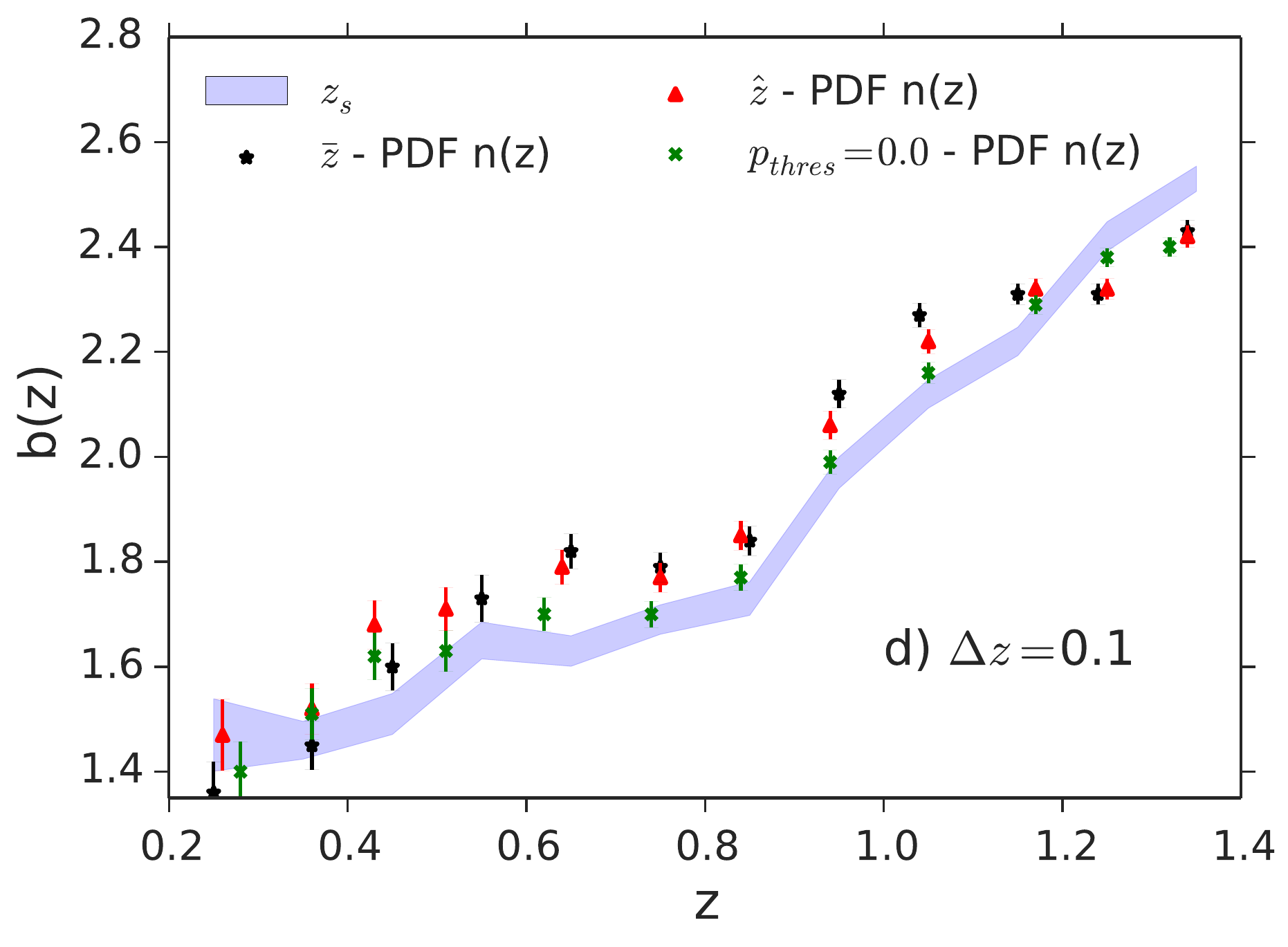}\\
\end{tabular}
\caption{\emph{Bias evolution for different redshift bin configurations:} The evolution with redshift of the linear galaxy bias when dividing the full sample into redshift bins. Results are shown both for the different bin widths and the different photometric redshift statistics, described in section \ref{sec:estimators}:  spectroscopic redshift results (blue shadow), mean photo-z (black star), mode photo-z (red triangle), and photo-z PDF weights (green cross). In each bin, the spectroscopic sample is different than the corresponding photo-z sample, thus we cannot directly compare them. The x-axis position is given by the mean redshift in each bin according to the $n(z)$, which is given by stacking the photo-z PDFs.}
\label{fig:fig9}
\end{center}
\end{figure*}

\begin{figure*}
\begin{center}
\begin{tabular}{cc}
\includegraphics[trim = 1cm 0cm 0cm 0.5cm, width=0.48\textwidth]{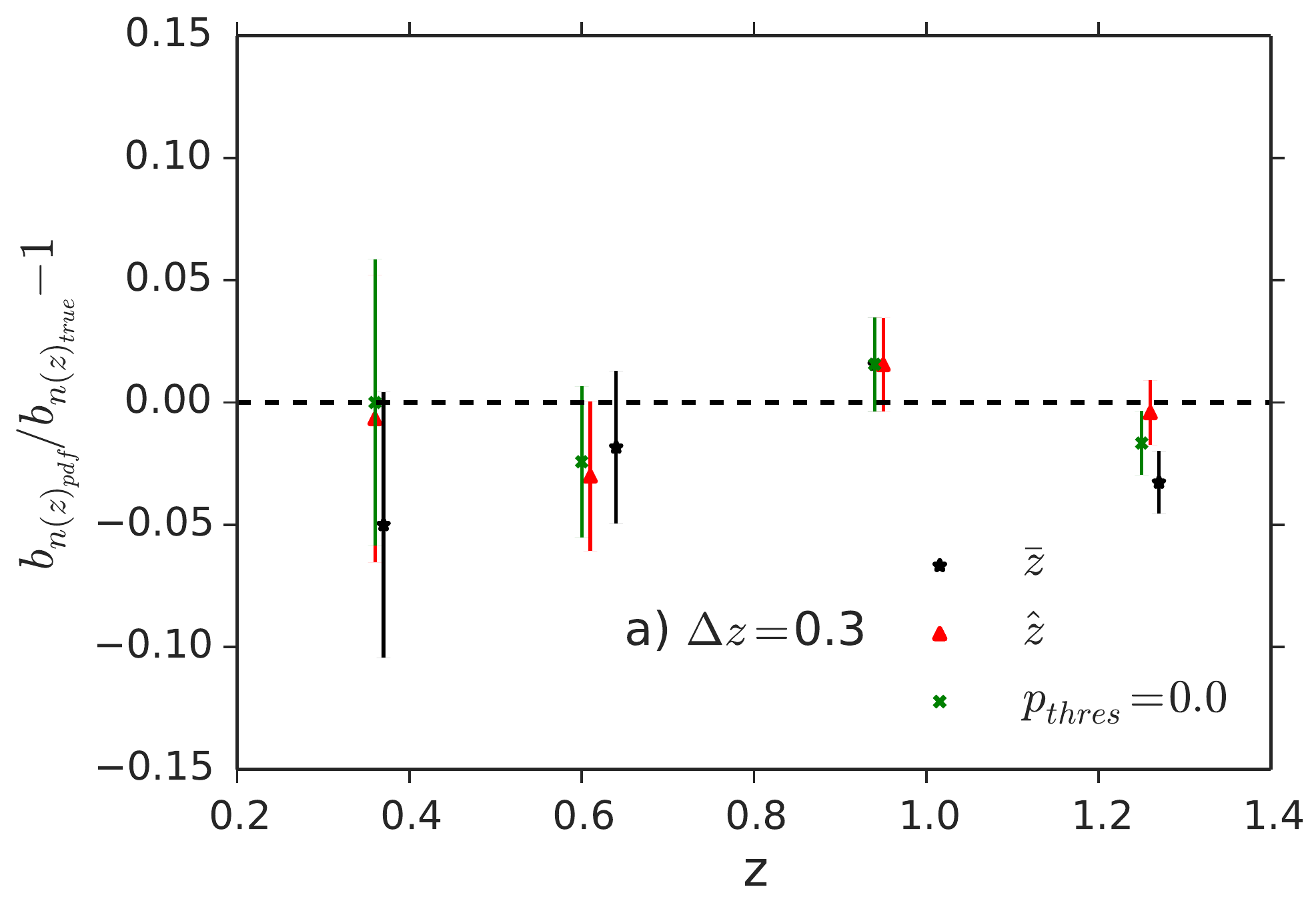} 
\includegraphics[trim = 0cm 0cm 1cm 0.5cm, width=0.48\textwidth]{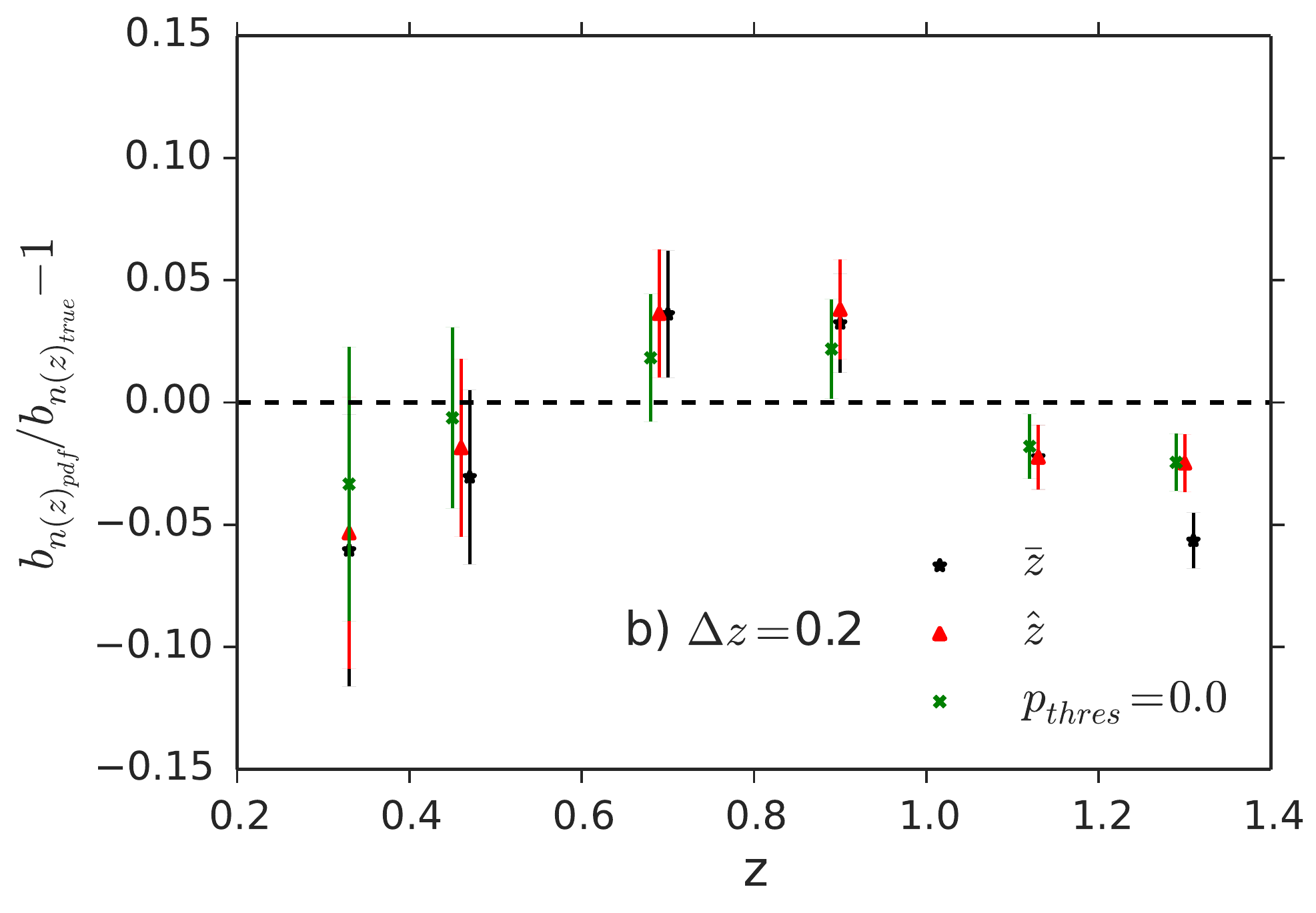} \\
\includegraphics[trim = 1cm 0cm 0cm 0.5cm, width=0.48\textwidth]{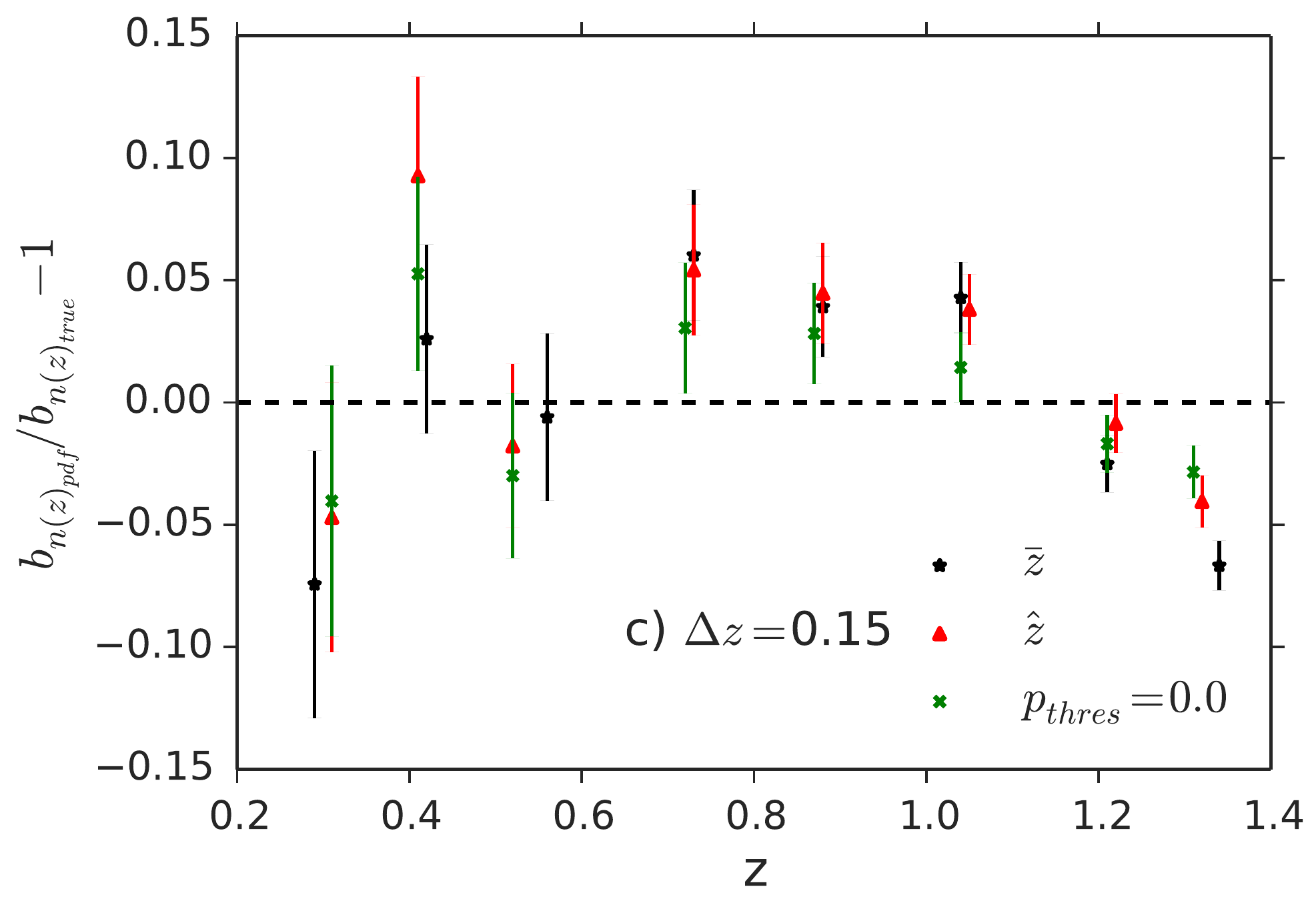}
\includegraphics[trim = 0cm 0cm 1cm 0.5cm, width=0.48\textwidth]{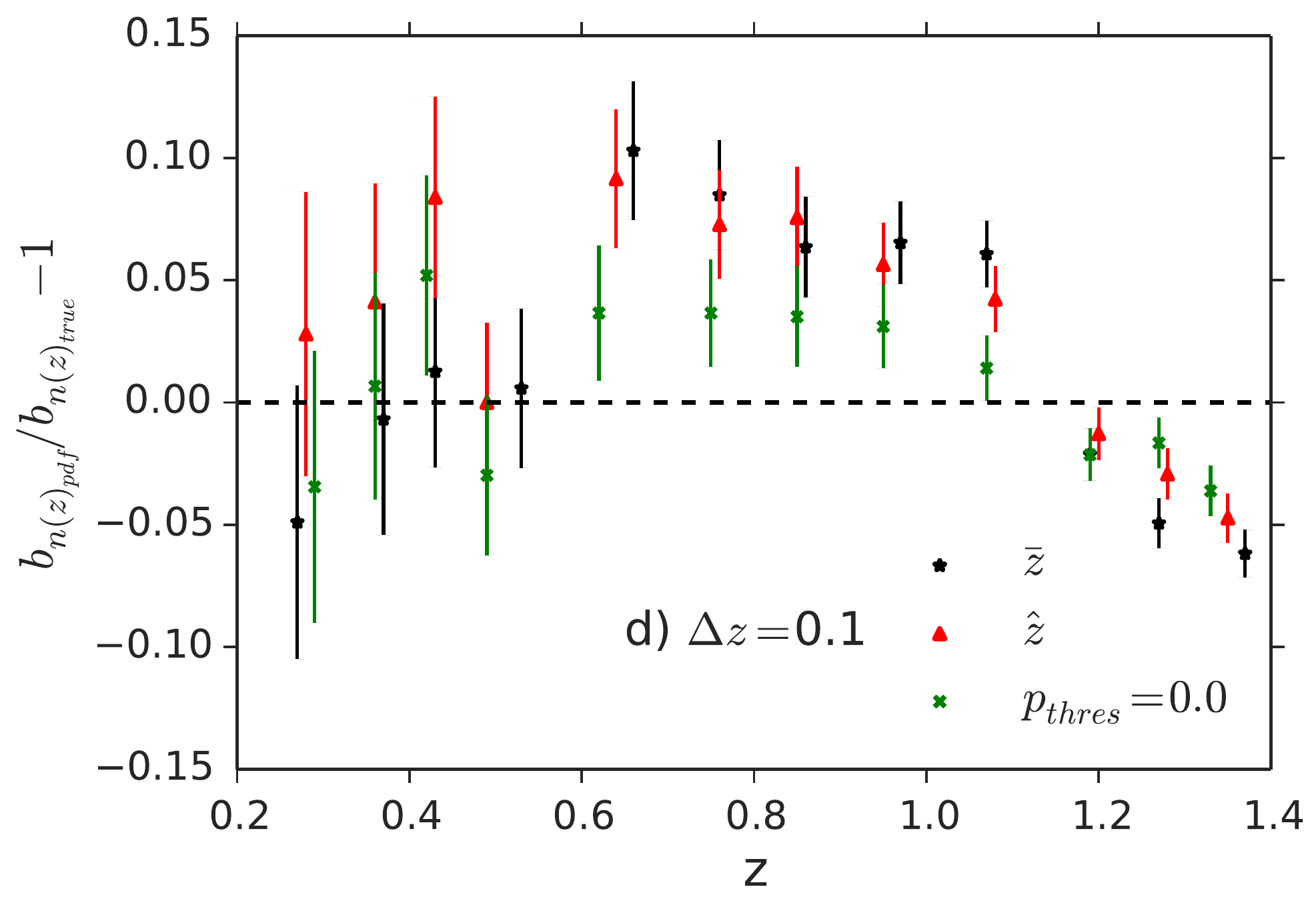}\\
\end{tabular}
\caption{\emph{Relative bias between PDF stacking and the true redshift distribution:} The relative differences in galaxy bias measurements when measuring the redshift distribution by using PDF stacking with respect to the bias measurements given by using the true redshift distribution. This computation is done for the three photo-z selection methods in each of the four redshift bin widths considered in this paper. Notice that for each method we are using the same photometric sample in each bin, but we are fitting the galaxy bias by using different redshift distributions.}
\label{fig:fig10}
\end{center}
\end{figure*}

We found that the evolution of the clustering amplitude with bin width evolves differently for the different estimators. 
We show in Figure \ref{fig:fig5b} the clustering amplitude evolution at $\theta=0.1$ degrees. For photometric redshift estimators, the clustering signal increases by about $50\%$ when increasing the numbers of bins by a factor of 4. Therefore, increasing the number of bins beyond the limit in which the bin width is comparable with the photo-z error is not an efficient process for any photo-z estimator, at least from the point of view of a clustering measurement, especially for bins with a width smaller than $\Delta z=0.15$, which is twice the mean photo-z error, as shown in Figure \ref{fig:fig3}.

\subsection{Bias measurement}\label{sec:res3}
\subsubsection{PDF redshift distributions}
We next evaluate how the selection of galaxies in radial shells when
using different photo-$z$ statistics affects the information on 
the linear galaxy bias $b_g$, as defined in Equation (\ref{eq:bias}). Fitting the galaxy bias, or any cosmological parameter, can help us to calibrate the 
effect of different photo-z statistics. We only used angular auto-correlation 
functions and we parametrized the galaxy bias by one parameter per redshift bin. 
For each redshift bin, we found the best fit bias, $b$, and its error by sampling a $\chi^2$ given by:
\begin{equation}
\chi^2(b)=\sum_{\theta,\theta^\prime}{(w_{obs}(\theta)-b^2w_{th}(\theta))C^{-1}_{\theta,\theta^\prime}(w_{obs}(\theta^\prime)-b^2w_{th}(\theta^\prime))}
\end{equation}
where the observed angular correlation $w_{obs}$ is given by Equation (\ref{eq:wobs}), the theoretical $b^2w_{th}$ is given by Equation (\ref{eq:wthetadetail}) and $C^{-1}_{\theta,\theta^\prime}$ is the inverse of the covariance matrix. 

The total redshift range considered is $0.2<z<1.4$, and all  bins have the same width for 
each configuration. The angular range considered was set to cover the co-moving coordinates range $10 h^{-1}Mpc<r<60 h^{-1}Mpc$, which corresponds to different angular ranges in each redshift bin. The minimum scale was selected by testing at which scale the linear growth model for the spatial correlation departs from a non-linear model. Notice that this is a conservative cut when compared with the cuts used in \cite{crocce15}. This corresponds to $\theta_{min}=0.8$ degrees at the lowest redshift and $\theta_{min}=0.19$ degrees at highest redshift bin. 

The comparison between the different photo-z selection methods is done by comparing each galaxy bias measurement with the true result, which is determined by using the true redshift distribution of the selected galaxies in order to do a fair comparison.  First, we show in Figure \ref{fig:fig9} the galaxy bias measurement made by using different photometric redshift statistics and the bias measurement done by selecting galaxies according to the spectroscopic redshift. Notice that the spectroscopic sample in each bin is different than the photometric samples considered, but since this is accounted for in the bias measurement, we can study how the photo-z statistic measurements compare with the spectroscopic one. 

In panel {\it a} of  Figure \ref{fig:fig9}, we show the evolution of galaxy bias for the broad $\Delta z=0.3$ bin configuration. We only show results for the true redshift, $z_s$, the mode redshift, $\hat{z}$, the mean redshift, $\bar{z}$, and the PDF weighted samples for clarity. We do not show here the results when applying photometric redshift quality cuts. The measurements are similar and the slightly different values for the different estimators are within the statistical error bars. This is reasonable as we are considering a broad redshift bins in this panel, and the differences between different photometric samples redshift distributions are thus small. 

The same trend is observed when considering $N_z=6$ redshift bins, as shown in panel {\it b} on Figure \ref{fig:fig9}. This case corresponds to bins with $\Delta z = 0.2$ width, which is larger than twice the photo-z dispersion, and, therefore, photometric redshift effects are still not the biggest issue. The evolution of linear galaxy bias with redshift resembles the results of \pcite{crocce15} for a MICECATv2.0 \cite{carretero15} sample, as the ratio $b_g(z=1.1)/b_g(z=0.3)$ is similar for both simulations.

We show in the lower left plot in Figure \ref{fig:fig9} the measurement of the bias for the different redshift estimators when considering $N_z=8$ redshift bins. In this case, we begin to observe bigger differences between the 
case when using photometric redshifts and the case when the bias was obtained by using spectroscopic redshifts, especially when compared to the previous cases that used larger bin widths. 

Finally, we show in the bottom right panel of Figure \ref{fig:fig9} the bias evolution when using $12$ bins of width $\Delta z=0.1$. The differences between the photo-z galaxy bias results and the spectroscopic bias measurement are larger than for the previous cases with broader bin configurations. The closest result to the spectroscopic results value of the bias is obtained when using the full PDF information $(p_{threshold}=0$), especially at intermediate redshifts.

We show in Figure \ref{fig:fig10} the measurement bias between the method that uses PDF stacking to estimate $n(z)$ and the true value, given by the $n(z)$ measured directly from the true redshifts from the simulation of the photometric sample. When considering full PDF information, we weighted the stacked PDFs and the corresponding true redshifts by the corresponding PDF weight.  We show in the top left panel the relative differences when considering $\Delta z=0.3$ for the three methods, finding small deviations with respect to the true results with minor differences between the different selection techniques. These observed differences exist because 
the PDF stacking technique is not perfectly reconstructing the true $n(z)$ of the population sample in the tomographic bins. In this case, the differences are small because the bin width is broad and photo-z systematics in the $n(z)$ are smaller.  

When we decrease the bin width to $\Delta z = 0.2$, the differences grow, as shown in the top right panel of Figure \ref{fig:fig10}. The three methods are still producing similar results, and, because the bins are still too broad, the relative bias is zero, within the error bars. When the configuration changes to bins with widths $\Delta z=0.15$, differences start to become more apparent and the PDF weighting method begins to differ from the single point estimate estimators. For the narrowest bin width configuration considered, $\Delta z = 0.1$, the differences at intermediate redshifts are larger than $5\%$ for single point estimators, whereas for PDF weighted galaxy samples, these differences are around $3\%$. We include a table in Appendix \ref{sec:ap2} that presents all galaxy bias measurements and the relative differences with the true results. 

\begin{figure}
\begin{center}
\includegraphics[trim = 0cm 0cm 0cm 0cm, width=0.48\textwidth]{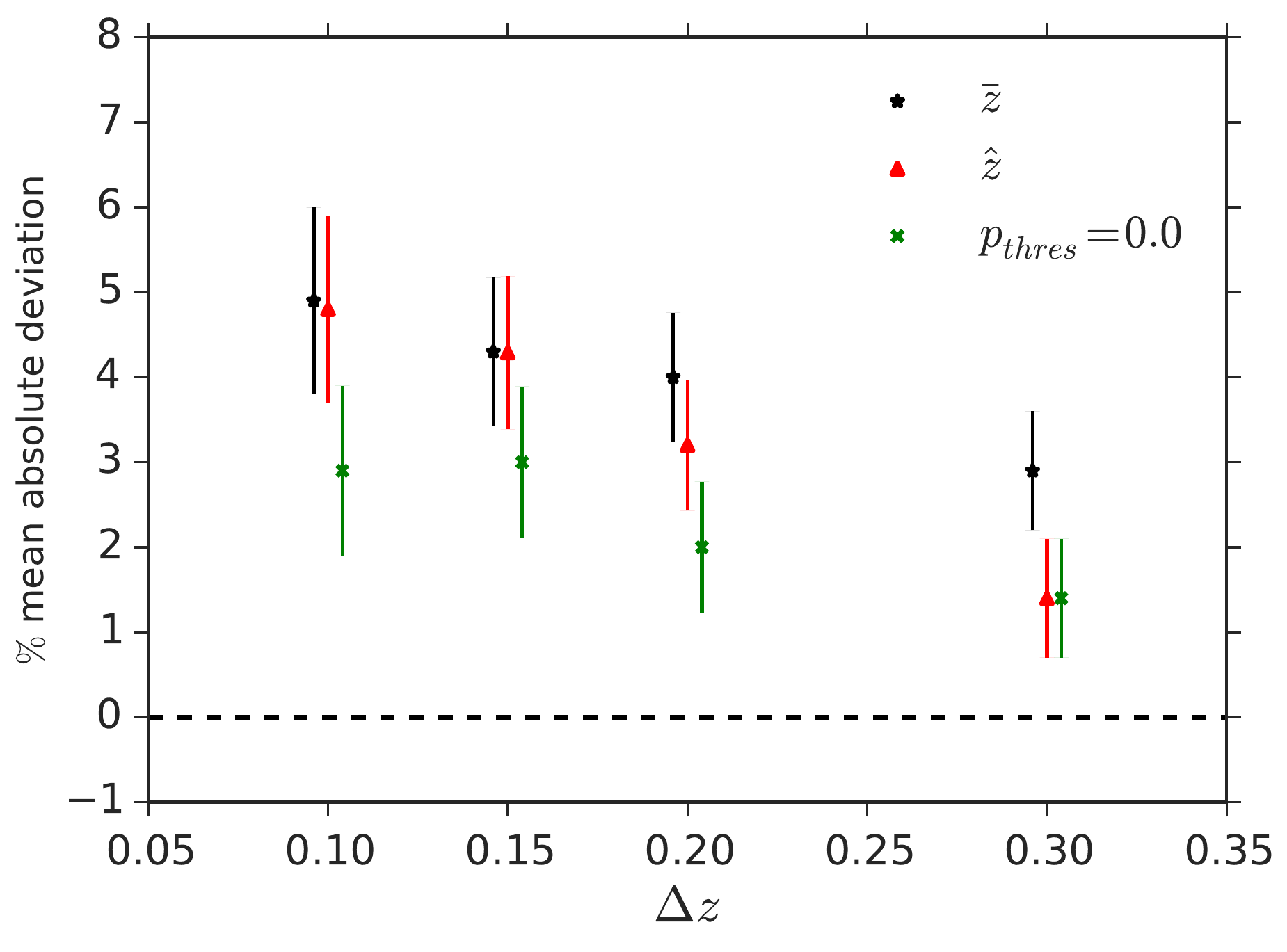}
\caption{Evolution with redshift bin width of the percentage deviation of the mean of the absolute difference with the true redshift result for the different photometric redshift statistics. We artificially shift the x-axis values in order to more clearly show the results from the different measures. Notice the accumulated measurement bias when using 
photo-z redshifts, which is smaller when using PDF weights in the clustering measurements, especially for the narrower bin configurations.}
\label{fig:fig13}
\end{center}
\end{figure}

In order to summarize and quantify these results, we show in Figure \ref{fig:fig13} the mean value of the mean absolute deviation between each selection method and the true result for each bin width.
We found that for the largest bin widths, the differences are around $1\%$ and are similar for the three photo-z selection statistics: mean, mode, and pdf weighting. However, for the narrower bins, the deviation when we consider summary statistics is around $5\%$, while it is $3\%$ when using the photo-z PDF galaxy weighting method.

\subsection{Reducing the redshift bin catalogue size}
Using full PDF information in galaxy clustering produces less biased measurements than point estimate photo-z methods, but it also increases the size of each redshift bin galaxy sample. We also studied how a Monte Carlo sampling of the PDF, in order to define a point estimate that encloses more of the PDF than the mean or the mode, or applying a threshold cut based on the amount of PDF in each bin compares to the full PDF inclusion method.

\subsubsection{Monte Carlo sampling redshift}
\begin{figure}
\begin{center}
\includegraphics[trim = 0cm 0cm 0cm 0cm, width=0.48\textwidth]{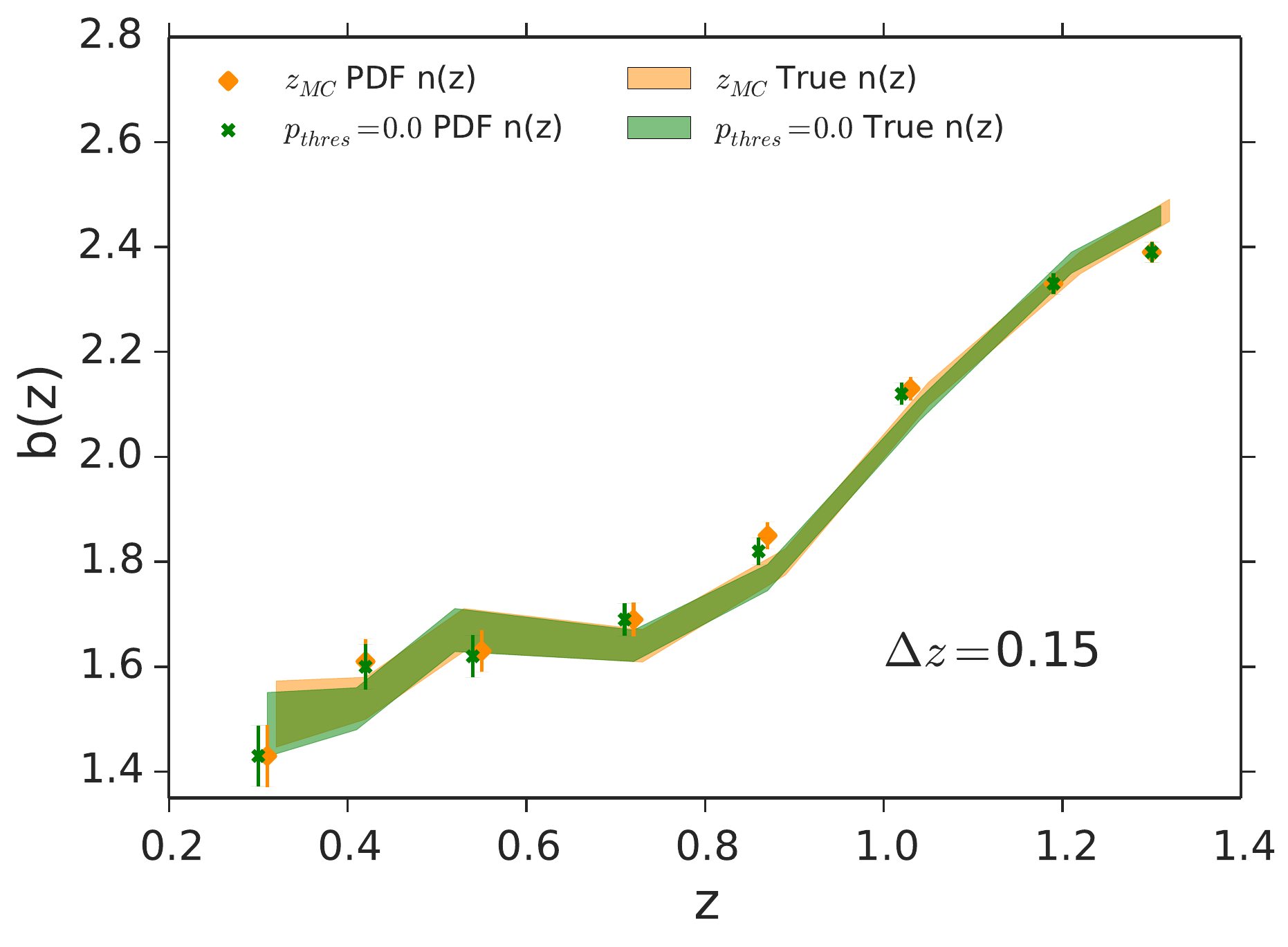}
\caption{\emph{Bias evolution in Monte Carlo sampling redshift shells of width $\Delta z = 0.15$:}  Bias measurement in $8$ redshift bins defined by the Monte Carlo sampling redshifts (orange). We consider the cases with $n(z)$ estimated by using PDF stacking and the true $n(z)$. We also compare the galaxy bias fitting with the PDF weights results from figure \ref{fig:fig9} (green). The standard performance of Monte Carlo redshifts is similar to the results given by PDF weights.}
\label{fig:fig11b}
\end{center}
\end{figure}

We extended the previous analysis to include a Monte Carlo sampling redshift, $z_{MC}$, which assigns a redshift value based on the cumulative distribution function for each galaxy. We make our previous galaxy bias measurement in the different redshift bins according to $z_{MC}$. In Figure \ref{fig:fig11b} we show the best fit galaxy bias for galaxies selected according to  $z_{MC}$ in $N_z=8$ redshift bins of $\Delta z = 0.15$.
We observe that in this case the results are similar to the results given by PDF weights (for example, the results from panel c of Figure \ref{fig:fig9}), both when using PDF sampling or the true redshift distributions. This is expected, as we are using the probabilistic information to determine the Monte Carlo sampling redshifts.

\subsubsection{Quality cuts}
\begin{figure}
\begin{center}
\includegraphics[trim = 0cm 0cm 0cm 0cm, width=0.48\textwidth]{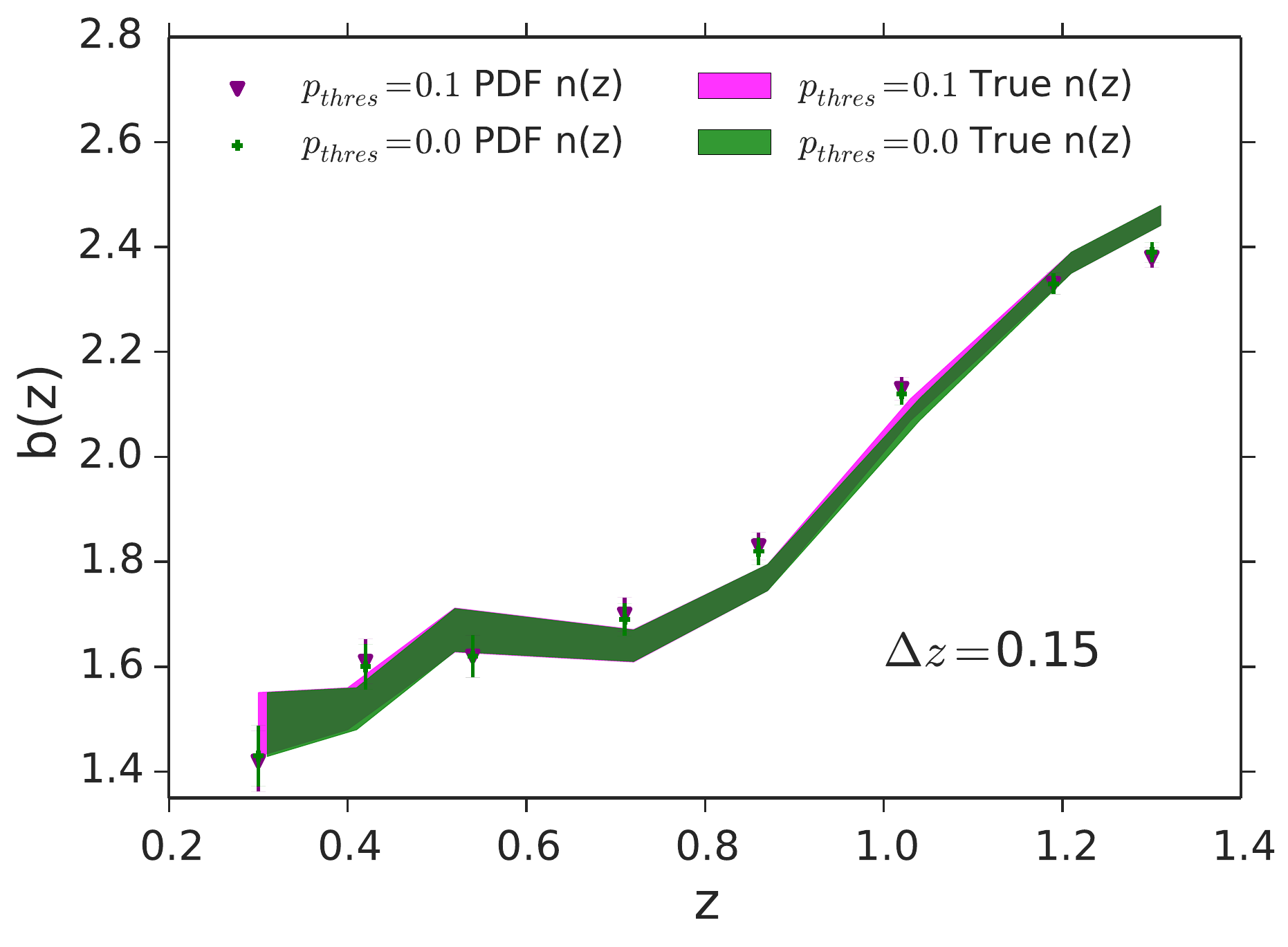}
\caption{\emph{PDF threshold cuts:} A comparison between the effect of applying a threshold cut on the selection process with photo-z PDF weights for 
the configuration in a redshift bin width of $\Delta z=0.15$. Both samples produce similar galaxy bias measurements.}
\label{fig:figqual}
\end{center}
\end{figure}

The effect of sparse PDFs with multiple peaks can introduce significant noise into our PDF weighting scheme. 
Although it is not the main interest of this paper, we considered a case in which we applied a threshold cut $f_z>(p_{threshold}=0.1)$ in 
order to select galaxies in the different bins. The effect is a combination of a quality cut and a cut on galaxies that are not in the bin but 
whose tails are inside the bin, which produces bigger catalogues in each tomographic bin. In Figure \ref{fig:figqual}, we present a comparison between a photo-z sample selected 
according to full photo-z PDFs for 
a configuration with bin width $\Delta z=0.15$ and a a sample selected by applying a threshold to the photo-z PDF weights of $f_z>0.1$.
We found that the results are similar, supporting the idea of applying threshold cuts to reduce the size of the galaxy 
density in each pixel in the map, although cuts to a sample have to be applied carefully in order to 
avoid introducing selection biases to the sample, see, e.g., \pcite{marti14}. We observe this effect in Figure \ref{fig:figqual}, as the true results for both samples are not exactly equivalent. A detailed study of using quality cuts from PDF information is outside the scope of this paper. 

\subsection{Systematics}
\subsubsection{Training set sample variance}
\label{sec:trainsel}
Since we use one galaxy sample from one particular pixel of the simulation for our photo-z training set, we wanted to demonstrate 
that the choice of this one pixel did not bias our results. As a result, we compared the spectroscopic $n(z)$ 
for the galaxies in our training sample with the mean spectroscopic $n(z)$ from ten randomly selected 
galaxy samples from the entire area, to demonstrate that our results were not dependent on the choice of a particular pixel. We are not, in this paper, 
exploring the more traditional concept of photometric redshift 'sample variance' as discussed, for example, in \pcite{cunha12}.

\begin{figure}
\begin{center}
\includegraphics[trim = 0cm 0cm 0cm 0cm, width=0.48\textwidth]{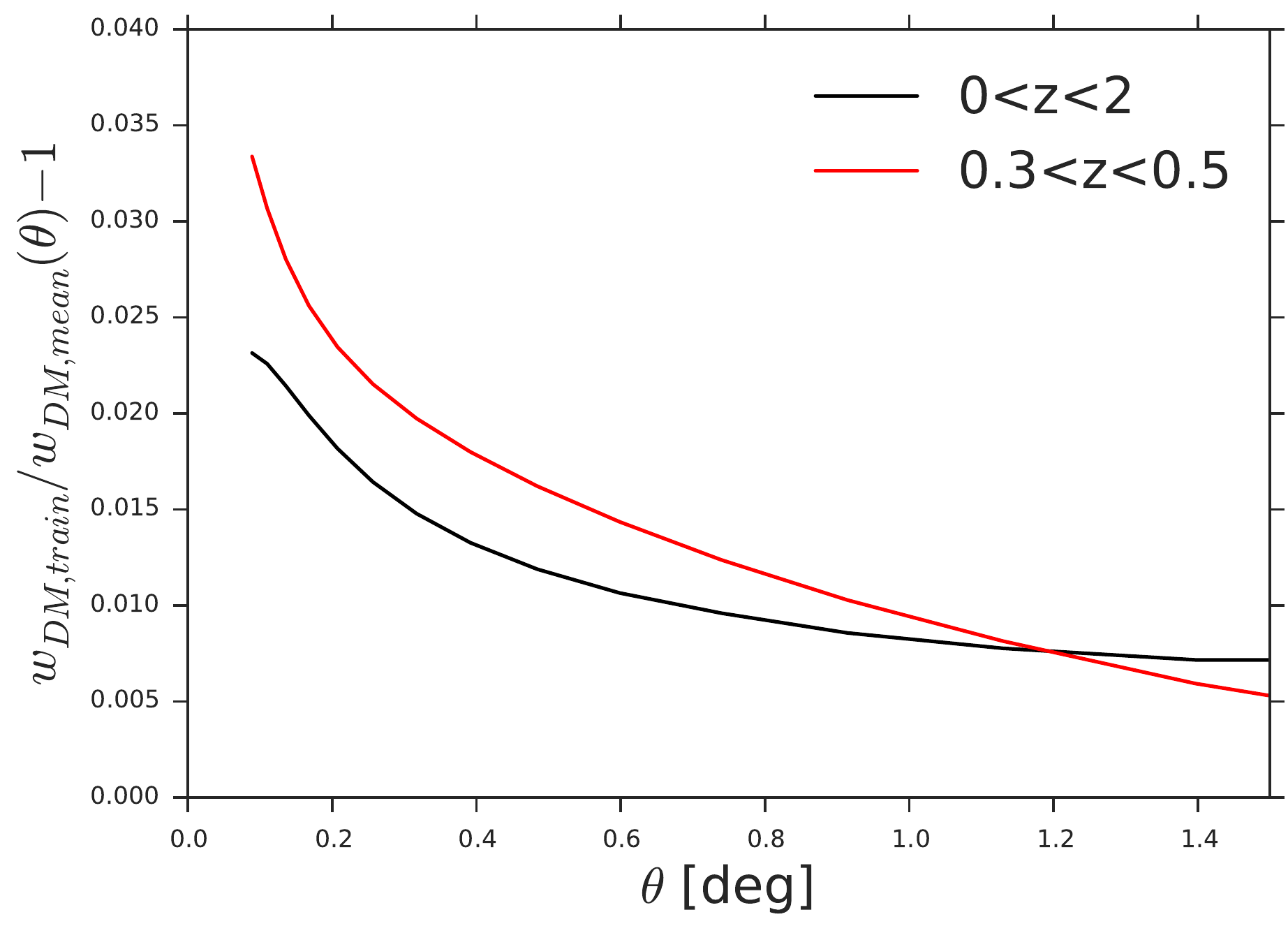}
\caption{\emph{Training set sample variance:} Relative differences between the amplitude of the theoretical dark matter angular correlations when using the spectroscopic $n(z)$ of the training set and the mean $n(z)$ of ten samples with the same number of spectroscopic objects as the training set but distributed across the catalogue area. We considered both the full redshift range $0<z<2$ (black) and a redshift bin $0.3<z<0.5$ (red). We find a relative bias in the angular range $0.1<\theta<1$ smaller than $2\%$, which propagates to an error smaller than $1\%$ on the galaxy bias, which is lower than the observed galaxy bias described in this paper.} 
\label{fig:fig15}
\end{center}
\end{figure}
We show in Figure {\ref{fig:fig15}} the relative difference between theoretical angular correlations, computed by using Equation (\ref{eq:wtheta}), when we used the training set $n(z)$ (red line in Figure \ref{fig:fig2}) and the mean $n(z)$ of ten different random samples extracted from the catalogue with the same number of galaxies as the training set, which is similar to the blue solid line in Figure \ref{fig:fig2}.  We found relative differences smaller than $2\%$ over the redshift range for all angles, $0.1<\theta<1$. This implies a relative difference smaller than $1\%$ for the galaxy bias, which is lower than the differences observed for our different photo-z statistics. 

\subsubsection{True redshift distribution reconstruction}
\begin{figure}
\begin{center}
\includegraphics[trim = 0cm 0cm 0cm 0cm, width=0.48\textwidth]{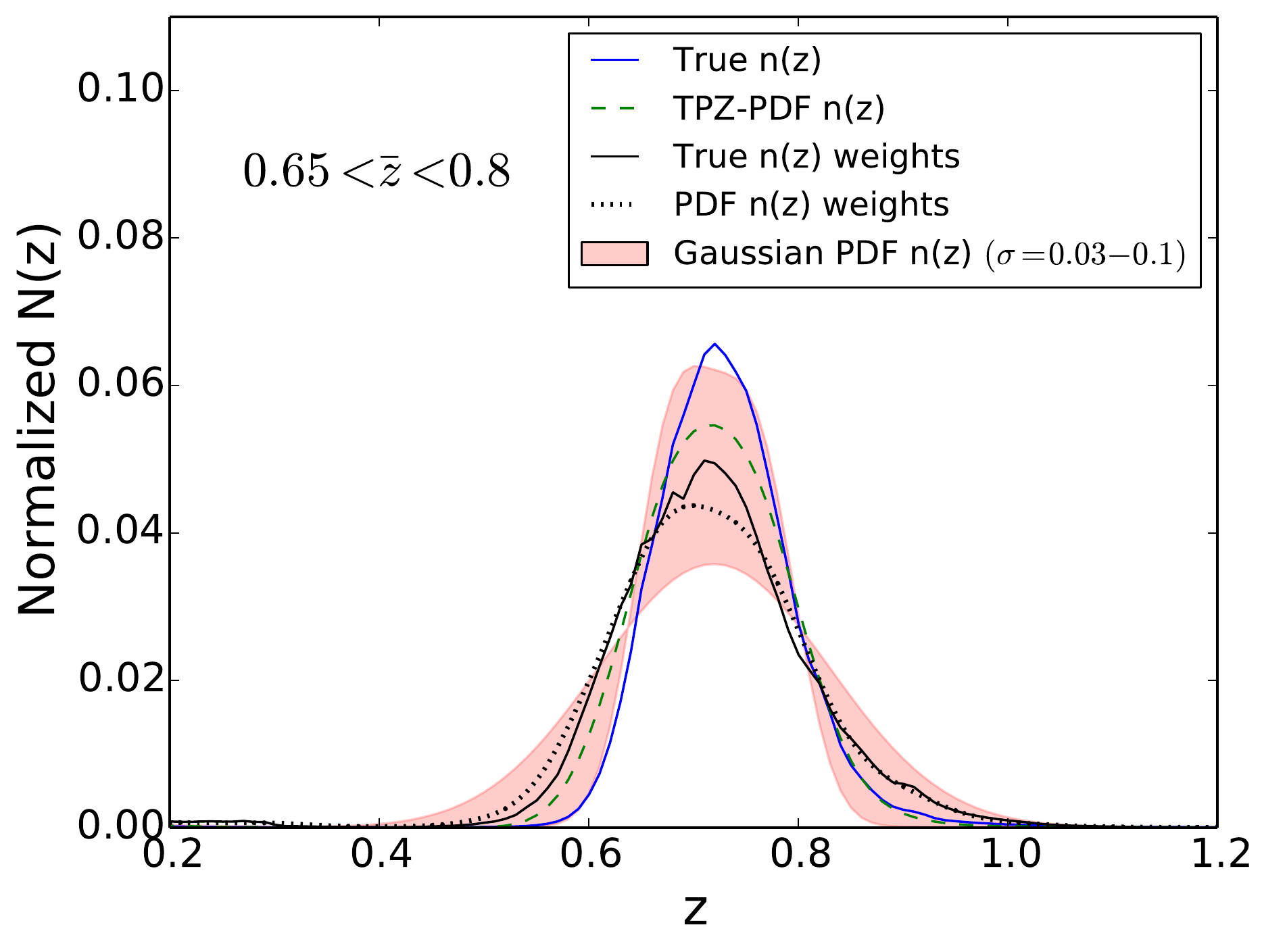}
\caption{\emph{Redshift distribution reconstruction:} A comparison between the PDF stacking $n(z)$ (dashed green) and the true redshift distribution (blue) obtained by selecting galaxies with mean photometric redshift in the redshift bin $0.65<z<0.8$. The red region covers the space between the $n(z)$ obtained by stacking gaussian PDFs for the galaxy sample with standard deviation within $\sigma_{gauss}=0.03-0.1$. The differences between redshift distributions are contained within the photo-z error. We compare with the redshift distribution of stacked weighted PDFs (solid black) and weighted true redshifts (dotted black) for the same redshift bin.}
\label{fig:fignzrecgauss}
\end{center}
\end{figure}

As observed in section \ref{sec:res3}, the main difference between the photometric redshift and the true redshift galaxy bias measurement is a result of the failure to recover the true redshift distribution, $n(z)$. As an example, in Figure \ref{fig:fignzrecgauss}, we show the difference between the true redshift distribution (blue line) and the PDF stacking PDF (green dashed line) for  galaxies selected with mean redshift within $0.65<z<0.8$.  
The red shadowed region shows the range between $n(z)$ created by stacking gaussian PDFs with standard deviations in the range $\sigma_{gauss}=0.03$ to $\sigma_{gauss}=0.1$. We see that  the difference between the measured $n(z)$ is within the accuracy of the photo-z catalogue, shown in Figure \ref{fig:fig3}. We also show, for comparison, the true redshift distribution from the weighted sample and the true weighted redshift distribution.  We explore this result in more detail in Appendix \ref{sec:ap3}, where we look at the differences between the $n(z)$ obtained from stacking the photo-z PDF of galaxies selected in redshift shells according to their mean redshift and the true distribution of the same sample when using both different bin configurations and different redshift ranges.

\section{Discussion}\label{sec:discussion}
In this paper, we have studied how the angular galaxy clustering obtained from photometric populations depends on the different statistical estimators used 
to assign galaxies to specific redshift bins. The primary estimators that we have considered are the mean and the mode of a galaxy's photo-$z$ PDF. 
We found differences between the different estimators, in part, since they produce different galaxy samples in each top hat photometric redshift bin. As a result, the clustering signal is different when using either the mean or the mode.

We also included the full PDF information in our clustering analysis by weighting each galaxy according to the 
integrated probability that the galaxy actually resided within each redshift bin. This clustering signal is smaller than the clustering signal from single point estimates samples. If we apply a threshold cut of $p_{threshold}=0.1$, the clustering amplitude increases. This is explained by the fact that when we consider a larger threshold the corresponding $n(z)$ is narrower than when considering all galaxies with non-negligible weights in a redshift bin. However, we also may be sampling different type of galaxies, since we are only selecting galaxies with higher probability to lie within the bin. 

We extended the comparison between the different photometric redshift statistical representations to a cosmological parameter estimation analysis by measuring the linear galaxy bias in different redshift bins. We find that, in general, the photo-z estimators produce similar results, especially when considering broad bins. We find that there is a relative bias with respect to the true galaxy bias results, since the PDF stacking redshift distributions in each bin differ from the true redshift distributions. For narrow bins, the selection method given by PDF weights produces less biased differences with respect to the true results. The mean deviation for a bin configuration with width $\Delta z=0.1$ is $3\%$ when using PDF weights, while it is $5\%$ when using summary or single point estimate statistics. Thus, the use of photo-z PDF weights to select galaxies in tomographic redshift bins in order to measure the galaxy clustering in a photometric survey produces more robust results than using single point estimates.   
We can use the methodology presented  in this paper to calibrate the effect of assigning galaxies to photo-z bins to ensure that the 
model parameters from simulations mimic the real data catalogues. This also applies to other photo-z methods that estimate PDFs  \cite{sanchez14,bonnet15,leistedt2015} as they will have similar behaviours.

Creating maps with PDF weights involves much larger data sets than catalogues of galaxies selected 
only by redshift. One way to reduce the amount of data is to apply a cautious PDF quality cut by using a threshold when considering PDF weights. We found similar results 
to the full PDF results, although any cut on a sample has to be carefully tested.
Another way to reduce the size of the catalogues, while still retaining a certain level of the PDF information, is by using Monte Carlo Sampling point estimates. 
We found that Monte Carlo Sampling estimators produce similar results to our PDF weight results.

The effect of choosing different photometric redshift training samples from the simulation on the calculation of the galaxy bias measurement is smaller than $1\%$, which is lower than the effects due to the different photo-z statistics used in this paper. Likewise, the differences between  stacking photo-z PDFs to compute the redshift distribution and the true redshift distribution are also within the photo-z errors.

\section{Conclusions}\label{sec:conclusion}
With photometric surveys, we can accumulate much larger galaxy samples in less time than with spectroscopic surveys. However, the lack of true redshifts restricts the quality of any 
radial information on such a survey, as photometric redshift are produced from multi-band imaging.  

Therefore, we need to set a statistical definition of a photometric redshift in order to identify which tomographic redshift bin contains a given galaxy. The search for an optimal definition 
is the main goal of this paper. The core analysis of this paper consisted of defining a new photo-z selection method that includes the full photo-z PDF information by weighting each galaxy in the redshift bin with the probability that the galaxy lies in that bin, and to compare this result with methods based on  single statistical estimates such as the mean or the mode of the photo-z PDF.

We found, using mock galaxy catalogues and a machine learning photo-z code, that if we use single point statistics, like the mode or the mean, there is an offset on the galaxy bias measurements. These bias measurements are obtained either by measuring photometric redshift distributions by stacking the individual photo-z PDFs or from the true redshift distribution of the same galaxies. This shift must be taken into account when considering similar large scale structure analyses that leverage galaxies drawn from photometric surveys. This corrective effect can be estimated by applying a similar method to measure the offset in the determination of the cosmological measurement of interest by using simulations in similar conditions to the expected photometric data. In our case, we used the galaxy bias as the metric to test different photo-z statistics, and we found that, for single point statistics, the cumulative deviation is a $5\%$ for a bin configuration with width $\Delta z= 0.1$.

Our results are closer to the ground truth if we weight the contribution of each galaxy to a photo-z bin according to the amount of their photo-z PDF in each redshift bin. This approach, on the other hand, produces a difference of $3\%$ in the $\Delta z=0.1$. Therefore, and especially for narrow photometric top hat bins, PDF weighting is more optimal than simply using summary statistic photometric redshifts.

\section*{Acknowledgments}
JA and JJT acknowledge support from U.S. Department of Energy Grant No, DE-SC0009932. ISN would like to thank the Spanish Ministry of Economy and Competitiveness (MINECO) for funding support through grant FPA2013-47986-C3-2-P. RJB acknowledges support from the National Science Foundation Grant No. AST-1313415. RJB has been supported in part by the Center for Advanced Studies at the University of Illinois. This work also used resources from the Extreme Science and Engineering Discovery Environment (XSEDE), which is supported by National Science Foundation grant number OCI-1053575. We want to acknowledge M. Becker and R. Wechsler for their helpful guidance in properly using the simulation catalogue. We thank all the useful discussions  with DES members and especially those with G. Bernstein, M. Crocce, E. Gazta\nn aga, W. Hartley, K. Honscheid, A. Kim, A. Ross, E. Sanchez and C. Sanchez. 

Funding for the DES Projects has been provided by the U.S. Department of Energy, the U.S. National Science Foundation, the Ministry of Science and Education of Spain, 
the Science and Technology Facilities Council of the United Kingdom, the Higher Education Funding Council for England, the National Center for Supercomputing 
Applications at the University of Illinois at Urbana-Champaign, the Kavli Institute of Cosmological Physics at the University of Chicago, 
the Center for Cosmology and Astro-Particle Physics at the Ohio State University,
the Mitchell Institute for Fundamental Physics and Astronomy at Texas A\&M University, Financiadora de Estudos e Projetos, 
Funda{\c c}{\~a}o Carlos Chagas Filho de Amparo {\`a} Pesquisa do Estado do Rio de Janeiro, Conselho Nacional de Desenvolvimento Cient{\'i}fico e Tecnol{\'o}gico and 
the Minist{\'e}rio da Ci{\^e}ncia, Tecnologia e Inova{\c c}{\~a}o, the Deutsche Forschungsgemeinschaft and the Collaborating Institutions in the Dark Energy Survey. 

The Collaborating Institutions are Argonne National Laboratory, the University of California at Santa Cruz, the University of Cambridge, Centro de Investigaciones Energ{\'e}ticas, 
Medioambientales y Tecnol{\'o}gicas-Madrid, the University of Chicago, University College London, the DES-Brazil Consortium, the University of Edinburgh, 
the Eidgen{\"o}ssische Technische Hochschule (ETH) Z{\"u}rich, 
Fermi National Accelerator Laboratory, the University of Illinois at Urbana-Champaign, the Institut de Ci{\`e}ncies de l'Espai (IEEC/CSIC), 
the Institut de F{\'i}sica d'Altes Energies, Lawrence Berkeley National Laboratory, the Ludwig-Maximilians Universit{\"a}t M{\"u}nchen and the associated Excellence Cluster Universe, 
the University of Michigan, the National Optical Astronomy Observatory, the University of Nottingham, The Ohio State University, the University of Pennsylvania, the University of Portsmouth, 
SLAC National Accelerator Laboratory, Stanford University, the University of Sussex, and Texas A\&M University.

The DES data management system is supported by the National Science Foundation under Grant Number AST-1138766.
The DES participants from Spanish institutions are partially supported by MINECO under grants AYA2012-39559, ESP2013-48274, FPA2013-47986, and Centro de Excelencia Severo Ochoa SEV-2012-0234.
Research leading to these results has received funding from the European Research Council under the European UnionÕs Seventh Framework Programme (FP7/2007-2013) including ERC grant agreements 
 240672, 291329, and 306478.

This paper has gone through internal review by the DES collaboration. The DES publication number for this article is DES-2015-0139. The Fermilab pre-print number is FERMILAB-PUB-15-571.

\appendix
\newpage

\section{Error distribution}\label{sec:ap1}
In order to check the robustness of the galaxy photo-z PDFs that we used in the paper, we estimated the distribution of the 
photometric standardized error of the photo-z BCC galaxies used in the paper, ($z_{phot}-z_{true}/\sigma$). 
The standard deviation, $\sigma$, is given by Equation (\ref{eq:sigmaphz}).
\begin{figure}
\begin{center}
\includegraphics[trim = 0cm 0cm 0cm 0cm, width=0.48\textwidth]{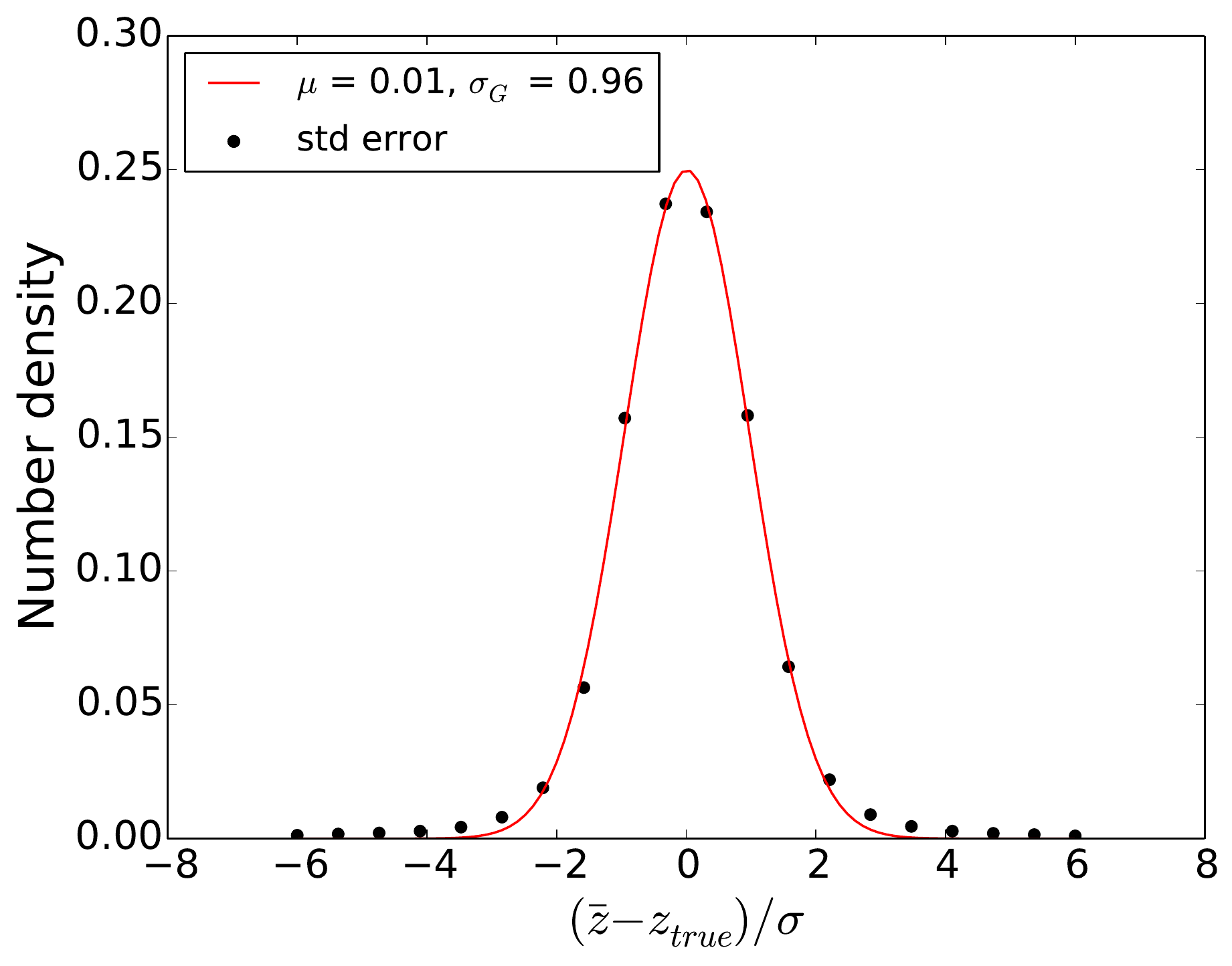}
\caption{\emph{Photometric standardized error for the mean}: The photometric standardized error computed from the mean of 
each individual galaxy's photo-z PDF compared to the best fit Gaussian, shown with the solid red line (mean $\mu$ and error $\sigma_G$).} 
\label{fig:figap1}
\end{center}
\end{figure}
In Figure \ref{fig:figap1}, we show the results using the mean redshifts. We observe that the simple error estimate is close to 
the unbiased estimate ($\mu=0$, $\sigma_G=1$). We can also consider the mode redshift, as shown in Figure \ref{fig:figap2}, where we see that 
the distribution of the modes tends to be more concentrated than the distribution of the means.
\begin{figure}
\begin{center}
\includegraphics[trim = 0cm 0cm 0cm 0cm, width=0.48\textwidth]{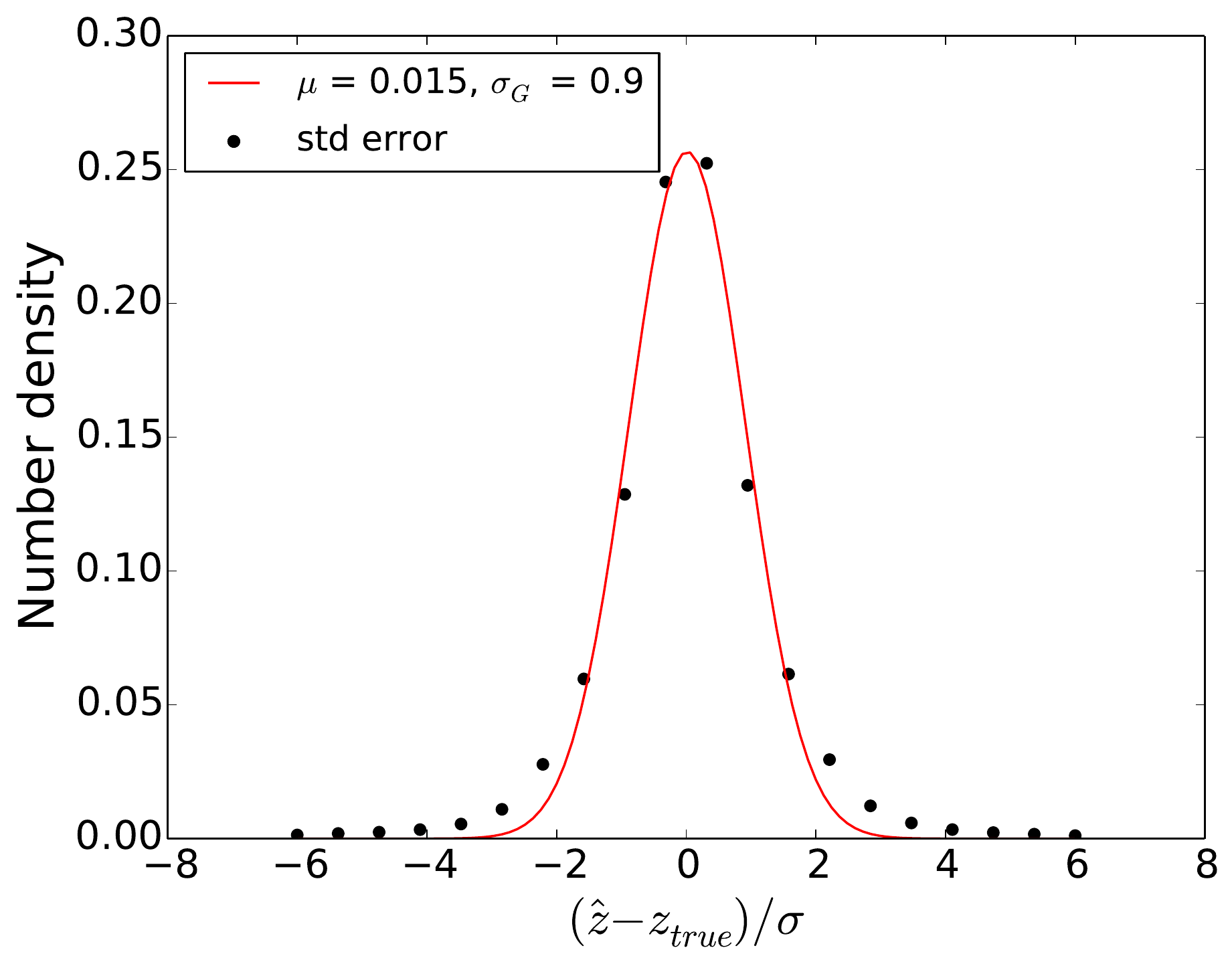}
\caption{\emph{Photometric standardized error for the mode}:  The photometric standardized error computed from the mode of 
each individual galaxy's photo-z PDF compared to the best fit Gaussian. As modes are defined by the peak of each PDF, the distribution tends 
to be more concentrated than the distribution of mean PDF values.} 
\label{fig:figap2}
\end{center}
\end{figure}

We also tested how photometric redshifts are distributed according to the confidence intervals 
by estimating the number of galaxies with photometric redshifts inside $1-\sigma$ and $2-\sigma$ levels, which is shown in Table \ref{table:tableap1}.
\begin{table}
{\center
\begin{tabular}{|c|c|c|}
\hline
 &$1-\sigma$ & $2-\sigma$ \\ 
\hline
$\bar{z}$&$70\%$&$93\%$  \\
\hline  
$\hat{z}$&$69\%$&$90\%$  \\
\hline  
\end{tabular}
\caption{Proportion of photometric redshifts inside $1-\sigma$ and $2-\sigma$ level confidence intervals for the mean, $\bar{z}$ and the mode, $\hat{z}$, photometric 
redshifts. In the ideal case, they are $68\%$ and $95\%$.}
\label{table:tableap1}
}
\end{table}
We see that the distribution of mean values in confidence intervals is close to the expected $68\%$ and $95\%$ distributions. When considering the mode, 
the values are more concentrated as we are considering the peaks of each individual PDF.

\section{Galaxy bias results}\label{sec:ap2}
In tables \ref{table:table1}, \ref{table:table1}, \ref{table:table1}, \ref{table:table1} we show the galaxy bias fits for the different bin configurations and the three photometric redshift methods: mean, mode, and PDF, used in this paper to 
select galaxies in tomographic redshift bins. In each case, we stack the galaxy photo-x PDFs to compute the redshift distribution. We also present the goodness of fit for each fit and the relative difference with the appropriate true measurement.

\begin{table*}
{\center
\resizebox{\textwidth}{!}{\begin{tabular}{|c|c|c|c|c|c|c|c|c|c|}
\cline{2-10}
\multicolumn{1}{c}{}&\multicolumn{3}{|c|}{$\bar{z}$} &\multicolumn{3}{|c|}{$\hat{z}$}& \multicolumn{3}{|c|}{$p(z)$} \\ 
\cline{2-10}
\multicolumn{1}{c}{}&\multicolumn{2}{|c|}{PDF-stacking}&\multicolumn{1}{|c|}{Relative difference} &\multicolumn{2}{|c|}{PDF-stacking}&\multicolumn{1}{|c|}{Relative difference} &\multicolumn{2}{|c|}{PDF-stacking}&\multicolumn{1}{|c|}{Relative difference} \\ 
\hline
Photo-z bin &Galaxy bias & $\chi^2$&Comparison True ($n(z)$) &Galaxy bias &$\chi^2$/dof &Comparison True ($n(z)$)&Galaxy bias &$\chi^2$/dof &Comparison True ($n(z)$) \\ 
\hline
$0.2<z<0.5$&$1.52\pm0.064$ &12.3/7  &$-0.05\pm0.057$ &$1.50\pm 0.061$ &12.1/7 &$-0.01\pm0.057$ &$1.52\pm0.063$ &11.2/7 & $0\pm0.06$\\
\hline  
$0.5<z<0.8$& $1.61\pm0.036 $  & 0.99/7 &$-0.018\pm 0.032$ &$1.61\pm 0.037$ &0.92/7 &$-0.03\pm0.031$ &$1.61\pm 0.036$ &0.58/7 &$-0.02\pm0.03$ \\
\hline  
$0.8<z<1.1$&$1.96\pm0.026 $   & 7.6/7 & $0.016\pm0.020$ &$1.97\pm0.027 $ &8.3/7 & $0.015\pm0.019$&$1.95\pm0.026$ &7.4/7&$0.016\pm0.019$  \\
\hline  
$1.1<z<1.4$&$2.37\pm0.024 $ & 5.71/7 & $-0.033\pm 0.015$ &$2.39\pm0.024$ &8.9/7 &$0\pm0.014$ &$2.37\pm0.022 $ &7/7 &$-0.017\pm0.013$\\
\hline
\end{tabular}}
\caption{Galaxy bias measurements for photometric samples selected according to  the mean, $\bar{z}$, mode, $\hat{z}$, or PDF weighted galaxies in 
redshift bins, given different bin configurations.}
\label{table:table1}
}
\end{table*}
\begin{table*}
{\center
\resizebox{\textwidth}{!}{\begin{tabular}{|c|c|c|c|c|c|c|c|c|c|}
\cline{2-10}
\multicolumn{1}{c}{}&\multicolumn{3}{|c|}{$\bar{z}$} &\multicolumn{3}{|c|}{$\hat{z}$}& \multicolumn{3}{|c|}{$p(z)$} \\ 
\cline{2-10}
\multicolumn{1}{c}{}&\multicolumn{2}{|c|}{PDF-stacking}&\multicolumn{1}{|c|}{Relative difference} &\multicolumn{2}{|c|}{PDF-stacking}&\multicolumn{1}{|c|}{Relative difference} &\multicolumn{2}{|c|}{PDF-stacking}&\multicolumn{1}{|c|}{Relative difference} \\ 
\hline
Photo-z bin &Galaxy bias & $\chi^2$&Comparison True (n(z)) &Galaxy bias &$\chi^2$/dof &Comparison True (n(z))&Galaxy bias &$\chi^2$/dof &Comparison True (n(z)) \\ 
\hline
$0.2<z<0.4$&$1.40\pm 0.058$  & 14.6/7 &$-0.06\pm0.055$ &$1.42\pm 0.059$ & 14.2/7& $-0.05\pm0.056$&$1.45\pm 0.059$ &12.9/7 &$-0.03\pm0.056$\\
\hline  
$0.4<z<0.6$& $1.59\pm 0.041$  & 4.1/7 &$-0.03\pm 0.035$ &$1.59\pm 0.041$ &4.4/7 &$-0.02\pm0.027$ &$1.59\pm 0.042$ &4.5/7 &$-0.01\pm0.037$ \\
\hline  
$0.6<z<0.8$& $1.72\pm 0.031$ & 2.6/8 &$0.04\pm0.026$ &$1.71\pm 0.031$ &2.8/8 &$0.04\pm0.027$ &$1.67\pm 0.03$ &2.9/8 & $0.02\pm0.026$\\
\hline  
$0.8<z<1.0$&$1.91\pm 0.027$  & 4/7 & $0.03\pm0.021$&$1.91\pm 0.027$ &3.7/7 & $0.04\pm0.02$&$1.87\pm 0.026$ & 4.3/7& $0.02\pm0.02$\\
\hline
$1.0<z<1.2$& $2.19\pm 0.022$ & 12/8 &$-0.02\pm0.013 $&$2.18\pm 0.022$ &11.1/8 &$-0.02\pm0.014$ &$2.19\pm 0.021$ &12.9/8 &$-0.02\pm0.013$ \\
\hline  
$1.2<z<1.4$& $2.34\pm 0.021$ & 3.9/8 &$-0.06\pm0.011$ &$2.36\pm 0.021$ &5.5/8 & $-0.02\pm0.012$&$2.39\pm 0.019$ &6.9/8 &$-0.02\pm0.011$\\
\hline
\end{tabular}}
\caption{Galaxy bias measurements for photometric samples selected according to  the mean, $\bar{z}$, mode, $\hat{z}$, or PDF weighted galaxies in 
redshift bins of width $\Delta z = 0.2$.}
\label{table:table2}
}
\end{table*}
\begin{table*}
{\center
\resizebox{\textwidth}{!}{\begin{tabular}{|c|c|c|c|c|c|c|c|c|c|}
\cline{2-10}
\multicolumn{1}{c}{}&\multicolumn{3}{|c|}{$\bar{z}$} &\multicolumn{3}{|c|}{$\hat{z}$}& \multicolumn{3}{|c|}{$p(z)$} \\ 
\cline{2-10}
\multicolumn{1}{c}{}&\multicolumn{2}{|c|}{PDF-stacking}&\multicolumn{1}{|c|}{Relative difference} &\multicolumn{2}{|c|}{PDF-stacking}&\multicolumn{1}{|c|}{Relative difference} &\multicolumn{2}{|c|}{PDF-stacking}&\multicolumn{1}{|c|}{Relative difference} \\ 
\hline
Photo-z bin &Galaxy bias & $\chi^2$&Comparison True ($n(z)$) &Galaxy bias &$\chi^2$/dof &Comparison True ($n(z)$)&Galaxy bias &$\chi^2$/dof &Comparison True ($n(z)$) \\ 
\hline
$0.2<z<0.35$&$1.37\pm0.058$ & 12.5/7 & $-0.07\pm0.056$ &$1.42\pm0.062 $ &12.7/7 & $-0.05\pm0.059$&$1.43\pm0.058 $ &12.6/7 & $-0.04\pm0.055$ \\
\hline  
$0.35<z<0.5$&$1.58\pm0.042$ & 8/8 & $0.026\pm0.038$&$1.65\pm0.042$ &7.8/8 & $0.09\pm 0.039$&$1.60\pm 0.043$ & 6.5/8&$-0.05\pm0.04$ \\
\hline  
$0.5<z<0.65$& $1.66\pm0.042$ &3.6/7 & $0.01\pm0.035$& $1.66\pm0.042$&  1.9/7& $-0.01\pm 0.035$ &$1.62\pm0.04 $ & 1.2/7& $-0.03\pm 0.033$ \\
\hline  
$0.65<z<0.8$&$1.76\pm0.034$&2.9/7  &$0.06\pm0.029$  &$1.75\pm 0.033$ & 2.7/7& $0.05\pm0.28$ &$1.69\pm 0.031$ &2.7/7 & $-0.03\pm 0.027$ \\
\hline
$0.8<z<0.95$& $1.86\pm0.027$ & 2.3/7 &$0.04\pm0.021$ & $1.87\pm 0.027$& 2.2/7&$0.04\pm0.021$ &$1.82\pm 0.026$ &/7 & $0.028\pm 0.021$ \\
\hline  
$0.95<z<1.1$&$2.19\pm0.024$ &16/8  &$0.043\pm0.016$ &$2.18\pm 0.023$ & 16.7/8&$0.04\pm0.016$ &$2.12\pm 0.021$ & 17.2/8& $0.014\pm 0.014$ \\
\hline
$1.1<z<1.25$& $2.33\pm0.021$ & 8/7 &$-0.025\pm0.013$ &$2.33\pm 0.021$ & 11.4/7& $0.01\pm0.013$&$2.33\pm 0.02$ & 8.5/7&$-0.02\pm 0.011$ \\
\hline  
$1.25<z<1.4$&$2.38\pm0.020$& 3.9/8 & $-0.07\pm0.011$&$2.37\pm 0.021$ & 4.2/8 & $-0.04\pm0.011$&$2.39\pm 0.019$ & 6.8/8 &$-0.03\pm 0.011$ \\
\hline
\end{tabular}}
\caption{Galaxy bias measurements for photometric samples selected according to  the mean, $\bar{z}$, mode, $\hat{z}$, or PDF weighted galaxies in 
$\Delta z=0.15$ redshift bins.}
\label{table:table3}
}
\end{table*}

\begin{table*}
{\center
\resizebox{\textwidth}{!}{\begin{tabular}{|c|c|c|c|c|c|c|c|c|c|}
\cline{2-10}
\multicolumn{1}{c}{}&\multicolumn{3}{|c|}{$\bar{z}$} &\multicolumn{3}{|c|}{$\hat{z}$}& \multicolumn{3}{|c|}{$p(z)$} \\ 
\cline{2-10}
\multicolumn{1}{c}{}&\multicolumn{2}{|c|}{PDF-stacking}&\multicolumn{1}{|c|}{Relative difference} &\multicolumn{2}{|c|}{PDF-stacking}&\multicolumn{1}{|c|}{Relative difference} &\multicolumn{2}{|c|}{PDF-stacking}&\multicolumn{1}{|c|}{Relative difference} \\ 
\hline
Photo-z bin &Galaxy bias & $\chi^2$&Comparison True ($n(z)$) &Galaxy bias &$\chi^2$/dof &Comparison True ($n(z)$)&Galaxy bias &$\chi^2$/dof &Comparison True ($n(z)$) \\ 
\hline
$0.2<z<0.3$&$1.36\pm 0.059$ &8.7/7  & $-0.05\pm 0.059$ &$1.47\pm0.068 $ &8.1/7 & $0.028\pm 0.068$&$1.4\pm 0.057$ & 8.2/7&$-0.03\pm 0.056$\\
\hline  
$0.3<z<0.4$&$1.45\pm0.046 $ & 6.2/8 & $-0.01\pm 0.044$ &$1.52\pm 0.048$ &5.7/8 &$0.04\pm 0.046$ &$1.51\pm 0.049$ &/8 &$0.01\pm 0.046$\\
\hline  
$0.4<z<0.5$& $1.60\pm 0.045$& 7.9/7 & $0.01\pm 0.04$ &$1.68\pm0.046 $ & 7.3/7&$0.08\pm 0.042$ &$1.62\pm0.045 $ &6.2/7 & $0.05\pm 0.041$\\
\hline  
$0.5<z<0.6$&$1.73 \pm 0.045$ & 3.8/7 & $0.01\pm 0.037$ & $1.71\pm 0.041$& 1.7/7 &$0 \pm 0.033$ & $1.63\pm0.039 $&1.4/7 & $-0.03\pm 0.033$\\
\hline
$0.6<z<0.7$& $ 1.82\pm 0.033$ & 2.5/8 & $0.1\pm 0.028$ &$1.79\pm 0.033$ & 2.7/8 & $0.09\pm 0.028$ &$1.70\pm0.032 $ & 1.6/8 & $0.04\pm 0.028$ \\
\hline  
$0.7<z<0.8$&$1.79 \pm 0.028 $ & 12.3/8 & $0.08\pm 0.025$ &$1.77\pm0.028 $ & 13.1/8& $0.07\pm 0.024$ &$1.70\pm0.025 $ & 8.7/8&$0.04\pm 0.022$ \\
\hline
$0.8<z<0.9$& $ 1.84\pm0.028 $ & 0.85/7 &$0.06\pm 0.023$ &$1.85\pm 0.028$ & 0.8/7& $0.076\pm 0.023$ &$1.77\pm0.025 $ &2/7 & $0.04\pm 0.021$\\
\hline  
$0.9<z<1.0$&$ 2.12\pm0.027 $ & 7.3/7 & $0.07\pm 0.019$ &$2.06\pm 0.027$ & 6.7/7&$0.06\pm 0.020$ &$1.99\pm0.023 $ & 9.9/7& $0.03\pm 0.017$\\
\hline
$1.0<z<1.1$& $ 2.27\pm 0.023$ & 12.2/8 & $0.06\pm0.015$&$2.22\pm 0.023$ & 14.2/8 &$0.04\pm 0.015$ &$2.16\pm0.02 $ &16.2/7 & $0.014\pm 0.013$ \\
\hline  
$1.1<z<1.2$&$ 2.31\pm 0.02$ & 12.6/7 & $-0.02\pm 0.012$ &$2.32\pm 0.02$ &16.7/7 & $-0.01\pm 0.012$&$2.29\pm 0.018$ & 9.4/7& $-0.02\pm 0.011$ \\
\hline
$1.2<z<1.3$&$ 2.31\pm 0.02$  & 3.5/8 & $-0.05\pm 0.011$ &$2.32\pm 0.02$ &4.6/8 & $0.03\pm 0.011$&$2.38\pm0.018 $ & 7.2/8& $-0.02\pm 0.010$\\
\hline  
$1.3<z<1.4$&$ 2.43\pm0.021 $ & 4.1/8 & $-0.06\pm 0.011$ &$2.42\pm 0.021$ &5.1/8 &$-0.05\pm 0.012$ & $2.40\pm 0.018$&6.8/8 &$-0.04\pm 0.01$\\
\hline
\end{tabular}}
\caption{Galaxy bias measurements for photometric samples selected according to  the mean, $\bar{z}$, mode, $\hat{z}$, or PDF weighted galaxies in 
redshift bins of width $\Delta z=0.1$.}
\label{table:table4}
}
\end{table*}

\section{True redshift distributions}\label{sec:ap3}
\begin{figure*}
\begin{center}
\includegraphics[trim = 0cm 0cm 0cm 0cm, width=0.50\textwidth]{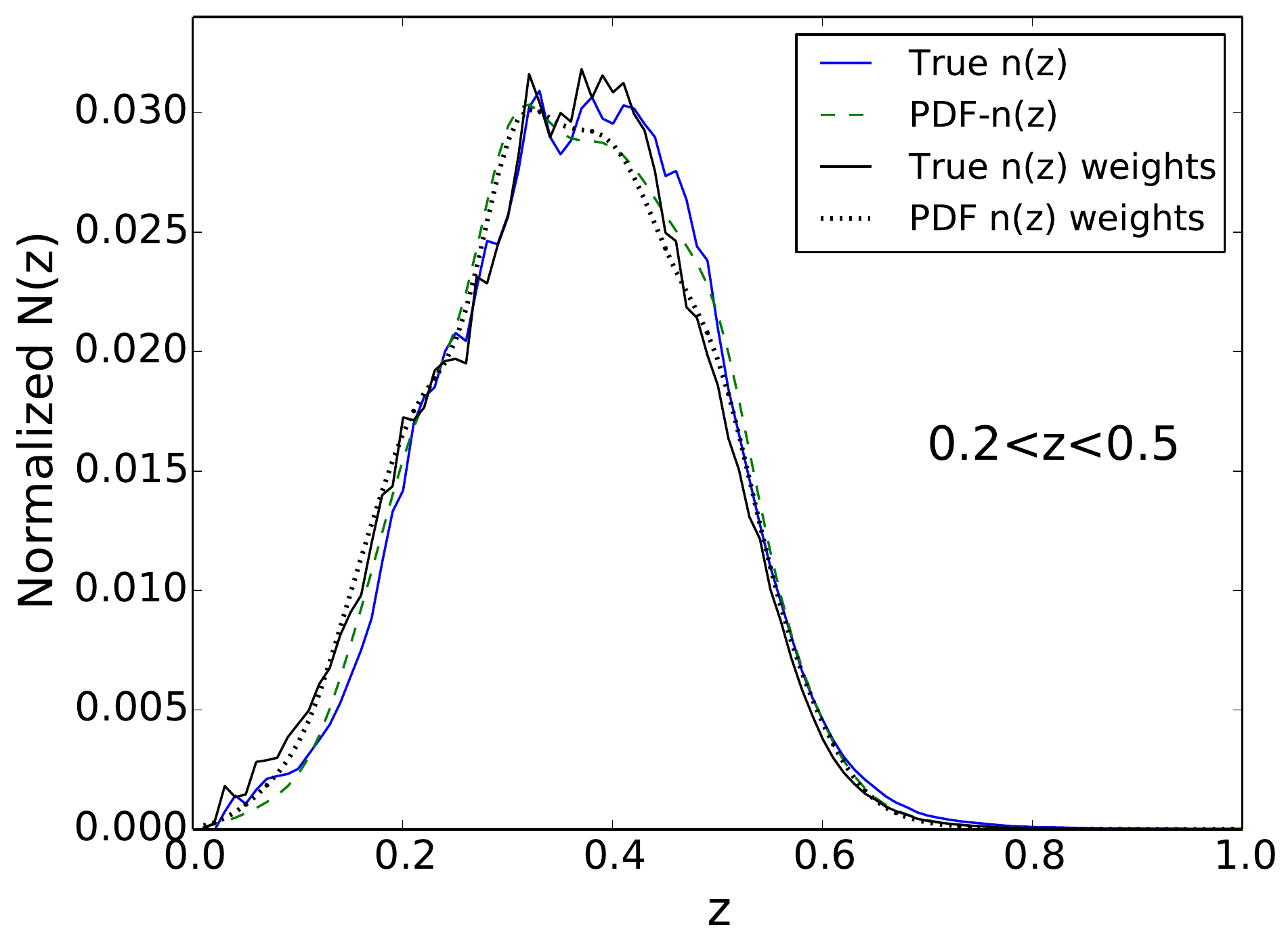}
\includegraphics[trim = 0cm 0cm 0cm 0cm, width=0.49\textwidth]{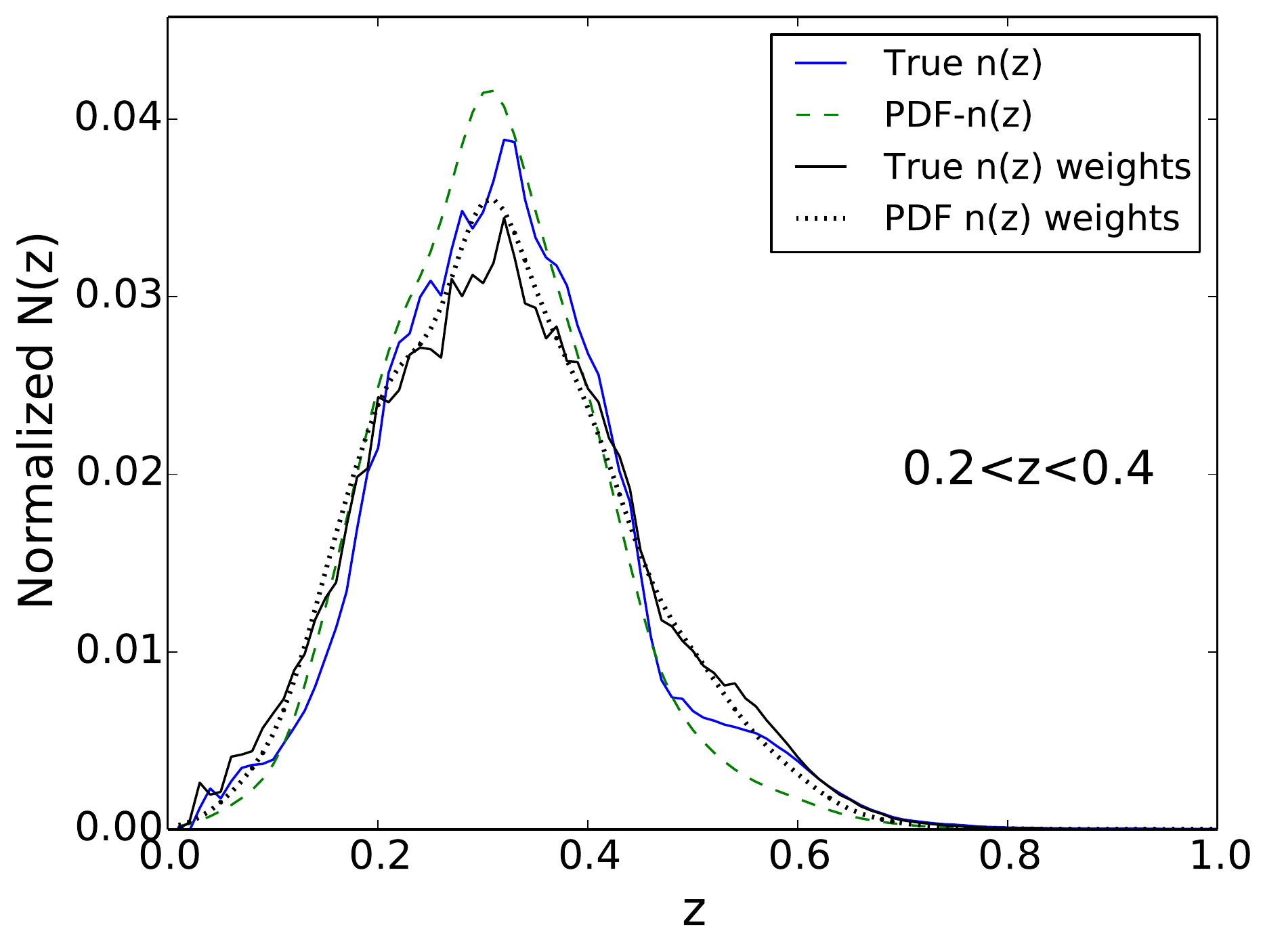}\\
\includegraphics[trim = 0cm 0cm 0cm 0cm, width=0.49\textwidth]{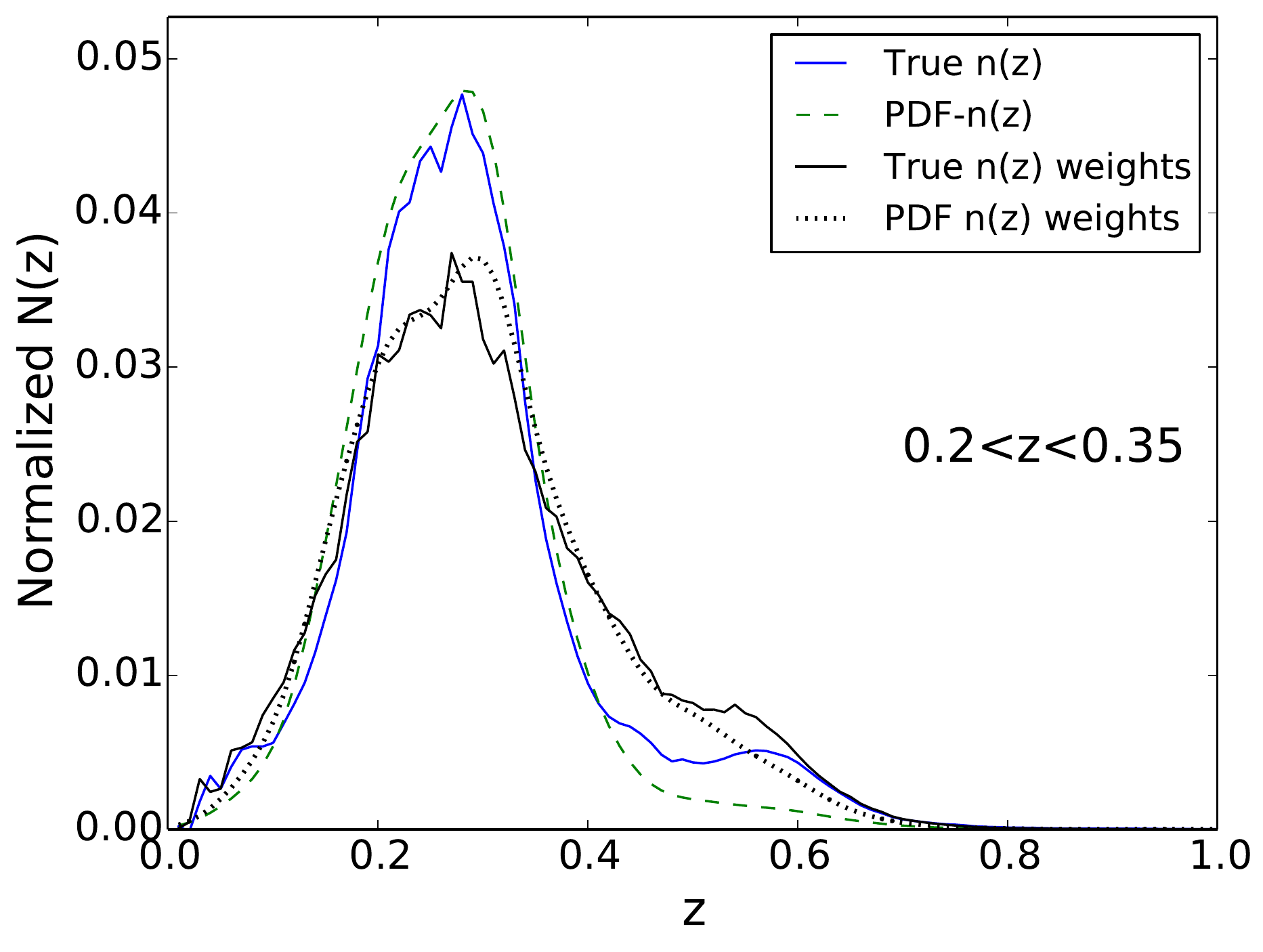}
\includegraphics[trim = 0cm 0cm 0cm 0cm, width=0.49\textwidth]{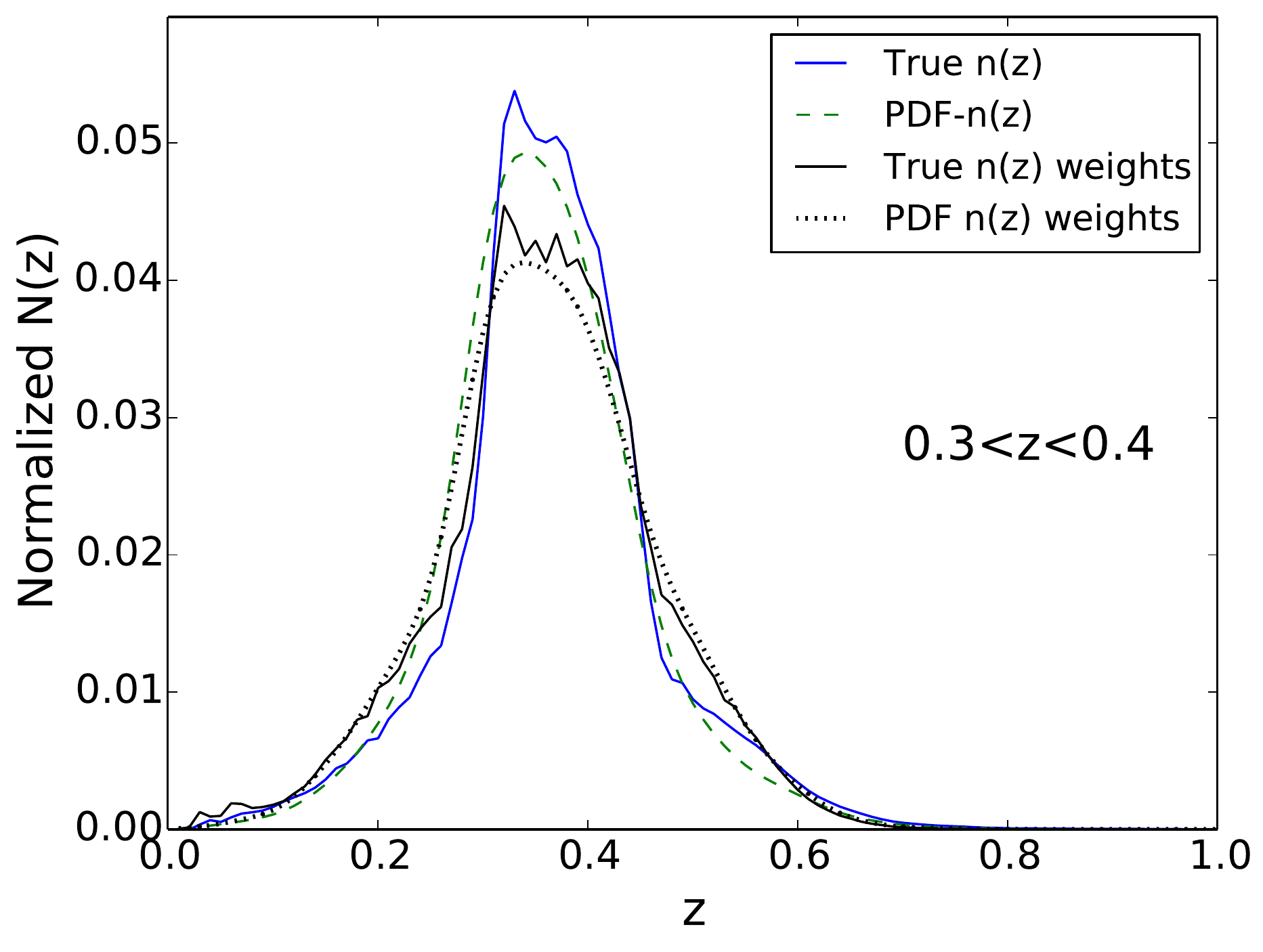}
\caption{\emph{Low redshift}:  A comparison of the redshift distributions for the true redshift distribution of galaxies selected according to mean photo-z redshift and the PDF stacking redshift distribution for the same galaxies over the lowest redshift range of the true galaxy sample. We also show the redshift distribution for galaxies selected according to the photo-z PDFs when stacking weighted PDFs (solid black) and true redshifts of weighted galaxies (dotted black).} 

\
\label{fig:figap3}
\end{center}
\end{figure*}

As shown in section \ref{sec:res3}, the galaxy bias measurements obtained from $n(z)$ given by PDF stacking are different than the true measurements. This is caused by 
the difference between the true redshift distribution of the photo-z galaxy sample and the PDF stacking $n(z)$. As an example, we show in Figure \ref{fig:figap3} the differences between the true redshift $n(z)$ and the PDF stacking $n(z)$ 
for galaxies selected according to the mean redshift for a given set of top hat redshift bins at low redshift. We note that the tails of the PDF stacking $n(z)$ are longer than

\begin{figure*}
\begin{center}
\includegraphics[trim = 0cm 0cm 0cm 0cm, width=0.50\textwidth]{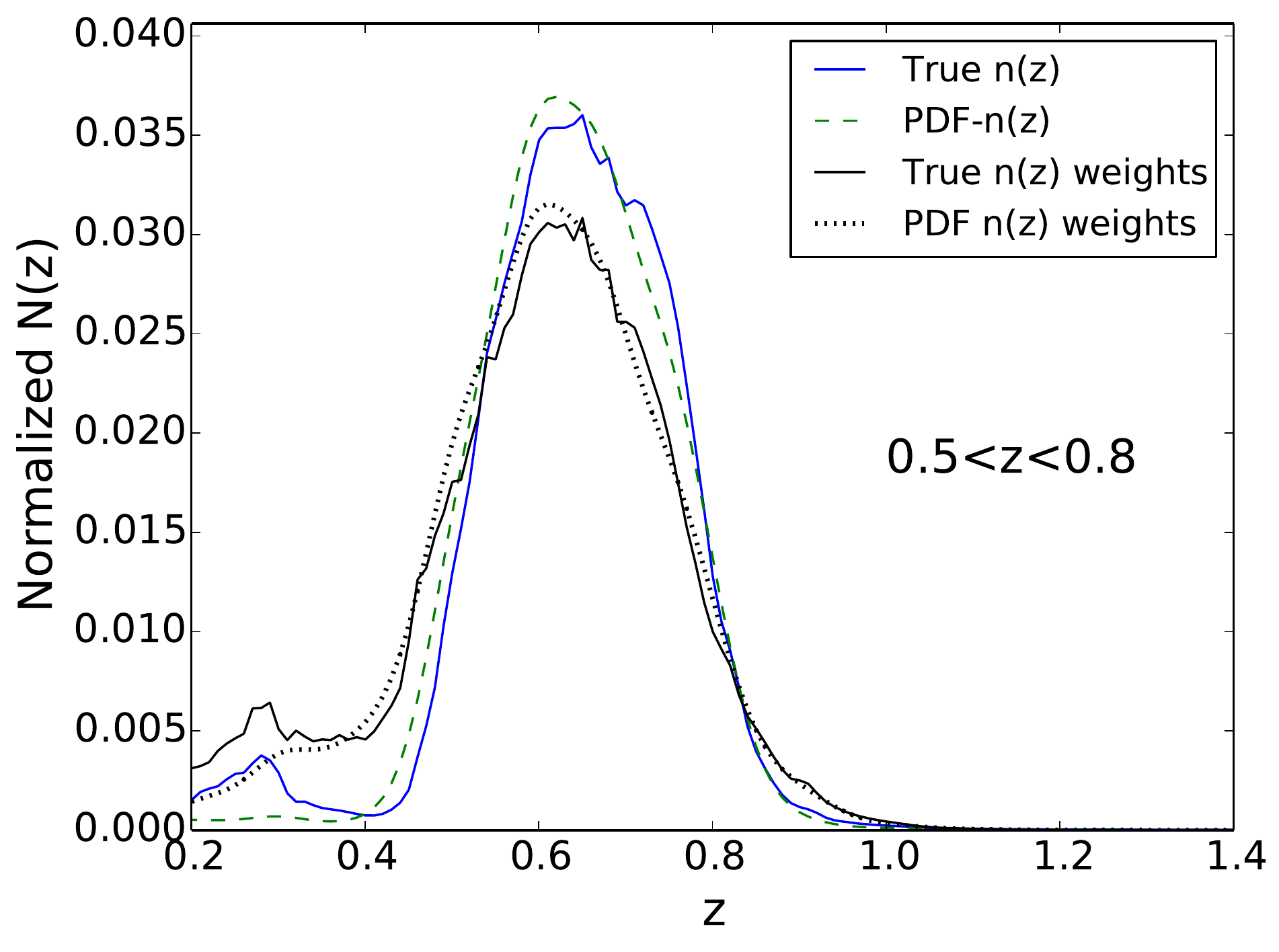}
\includegraphics[trim = 0cm 0cm 0cm 0cm, width=0.49\textwidth]{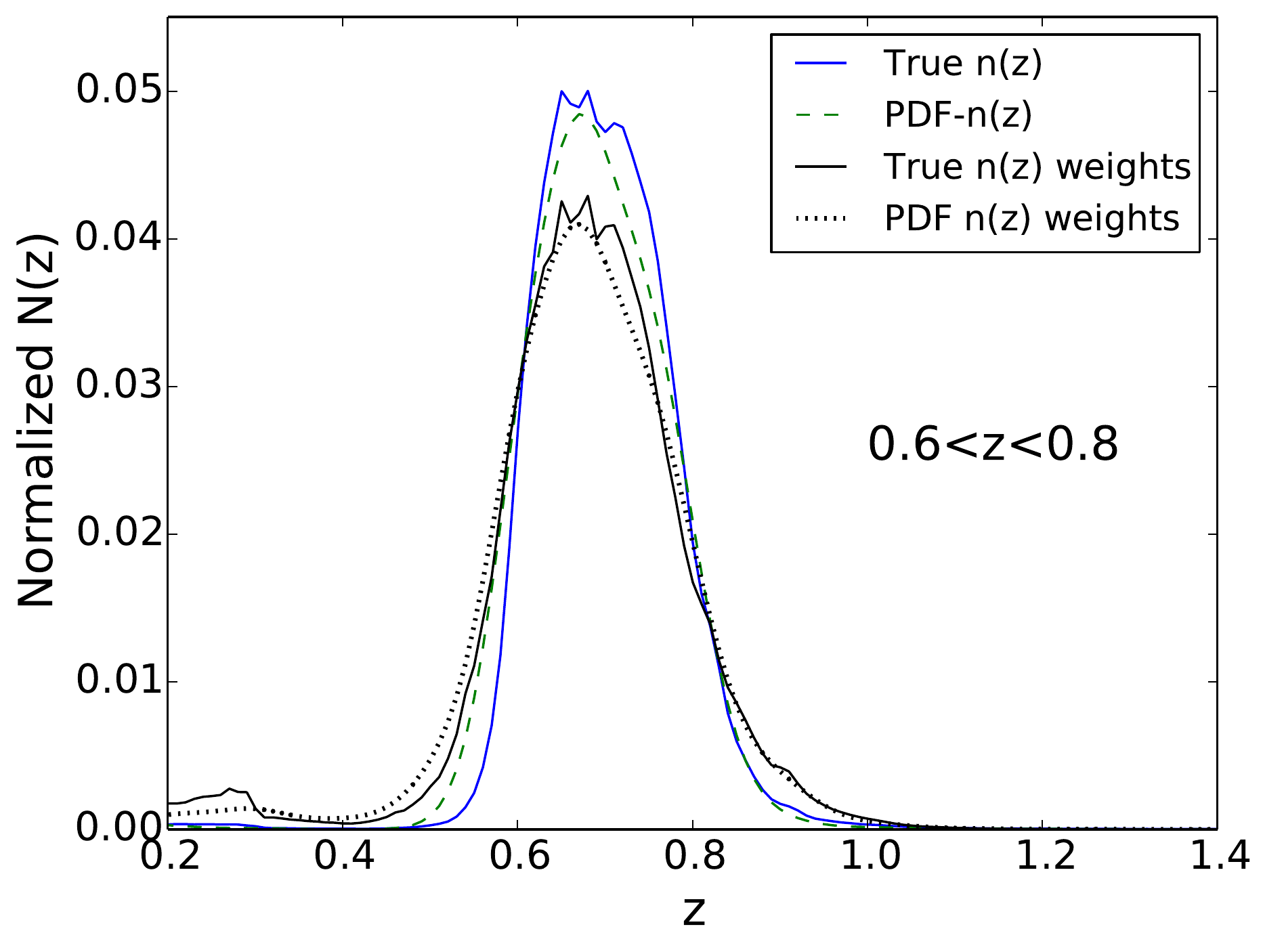}\\
\includegraphics[trim = 0cm 0cm 0cm 0cm, width=0.49\textwidth]{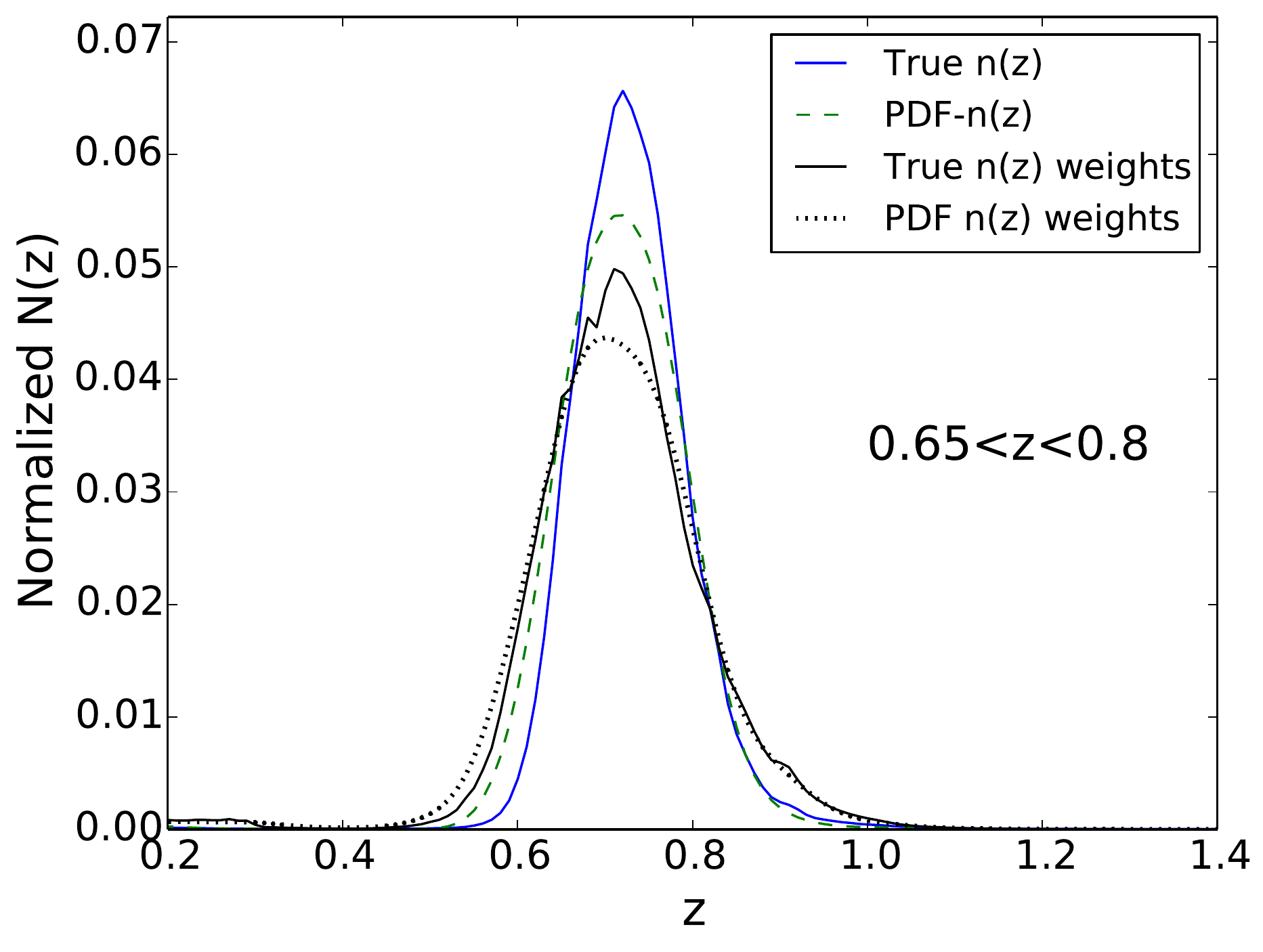}
\includegraphics[trim = 0cm 0cm 0cm 0cm, width=0.49\textwidth]{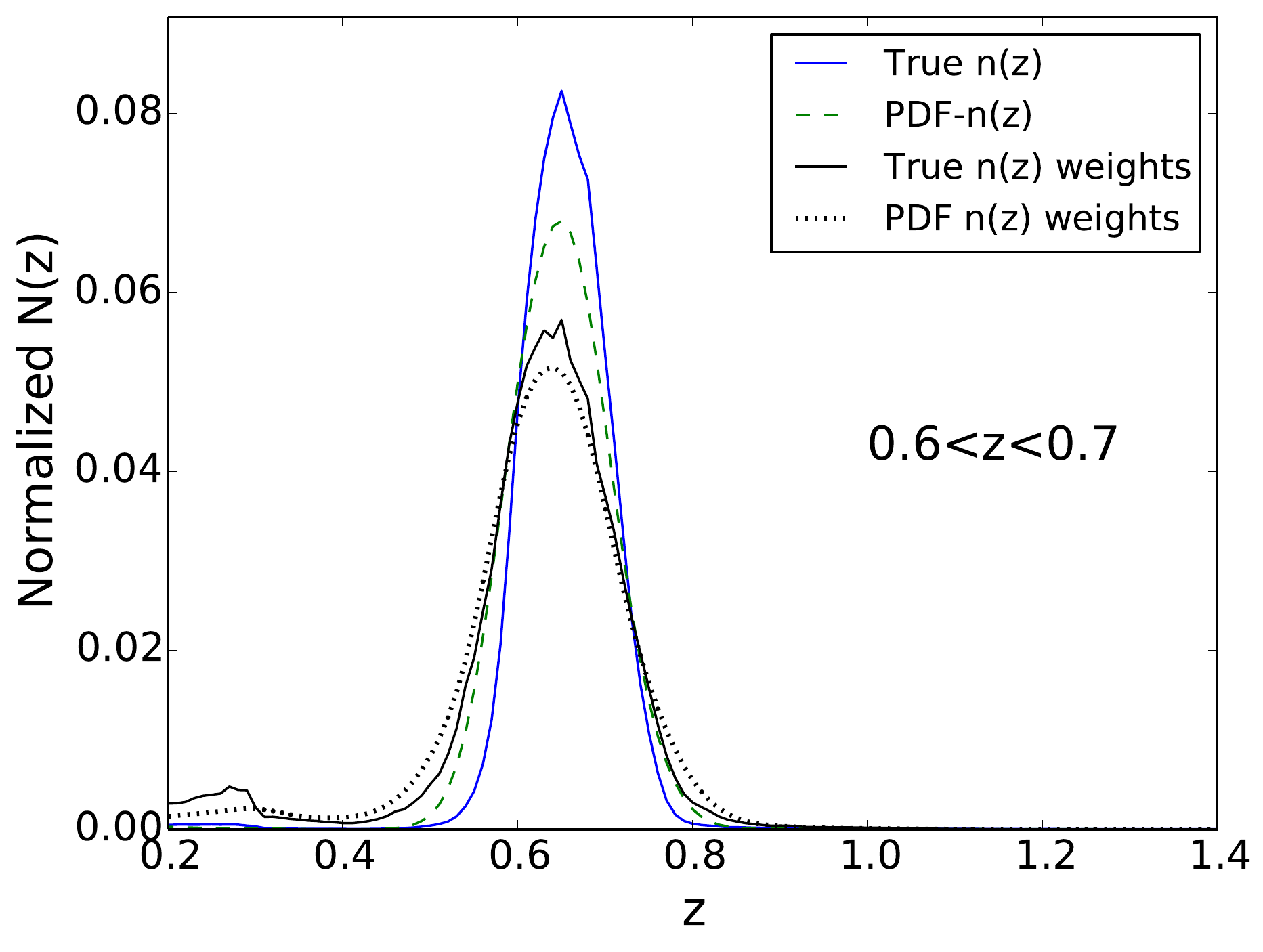}
\caption{\emph{Intermediate redshift}: A comparison between the true redshift distribution of galaxies selected according to mean photo-z redshift and the PDF stacking 
redshift distribution for the same galaxies over the intermediate redshift range of the true galaxy sample. The redshift distributions of galaxies selected according to PDF weights when stacking PDFs (solid black) or true redshifts of weighted galaxies (dotted black) are also displayed.} 

\
\label{fig:figap4}
\end{center}
\end{figure*}

\begin{figure*}
\begin{center}
\includegraphics[trim = 0cm 0cm 0cm 0cm, width=0.49\textwidth]{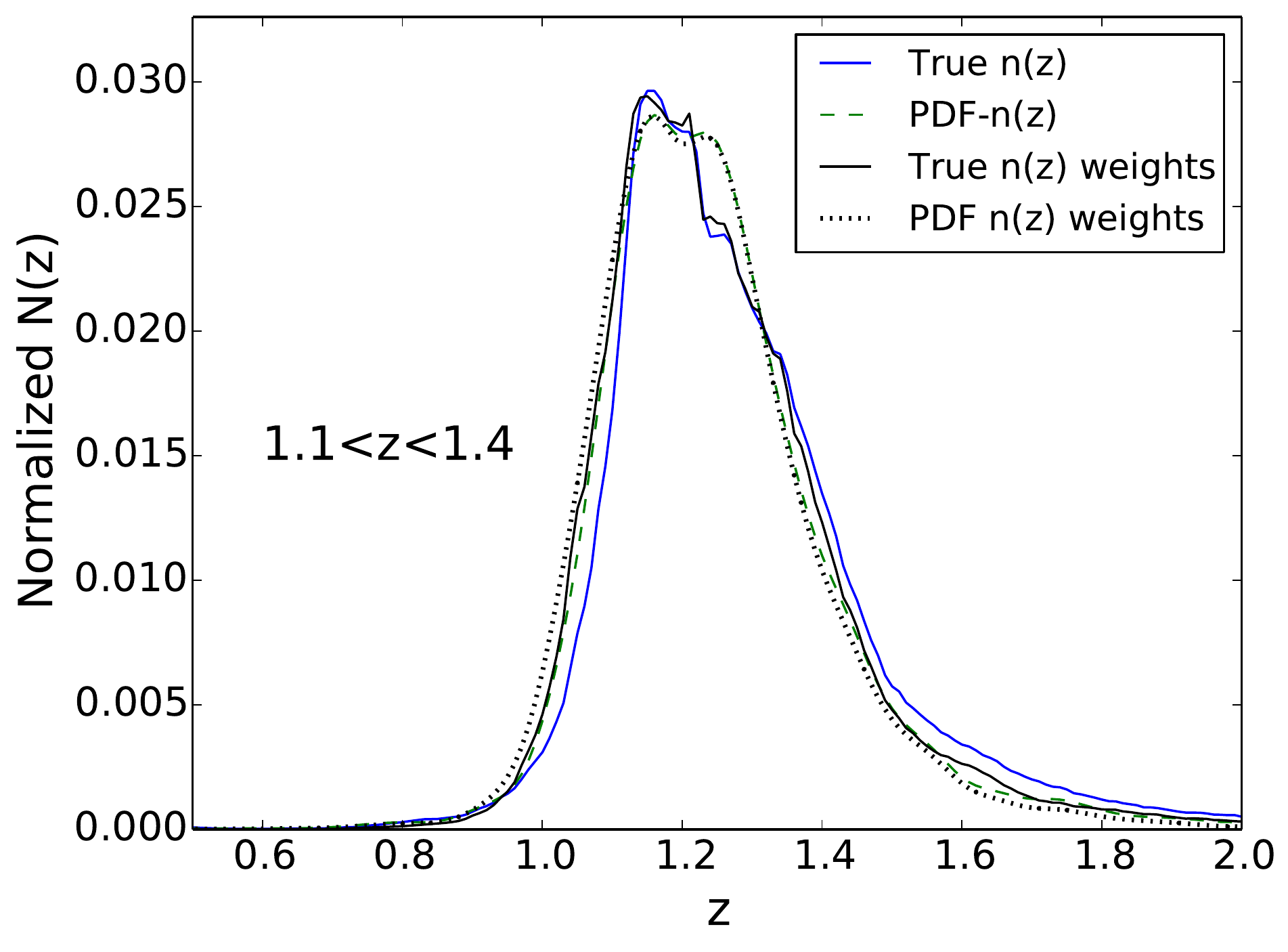}
\includegraphics[trim = 0cm 0cm 0cm 0cm, width=0.49\textwidth]{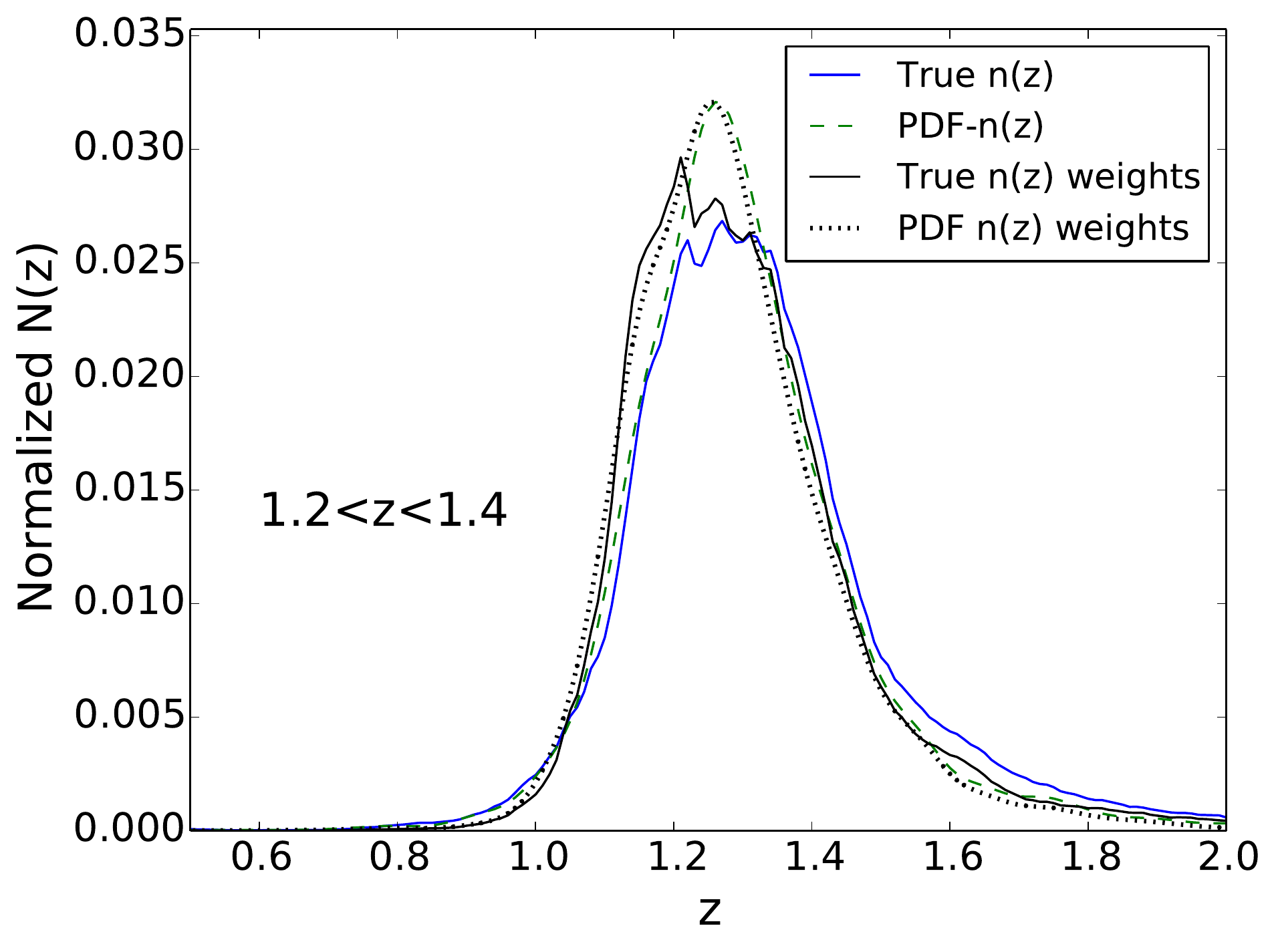}\\
\includegraphics[trim = 0cm 0cm 0cm 0cm, width=0.49\textwidth]{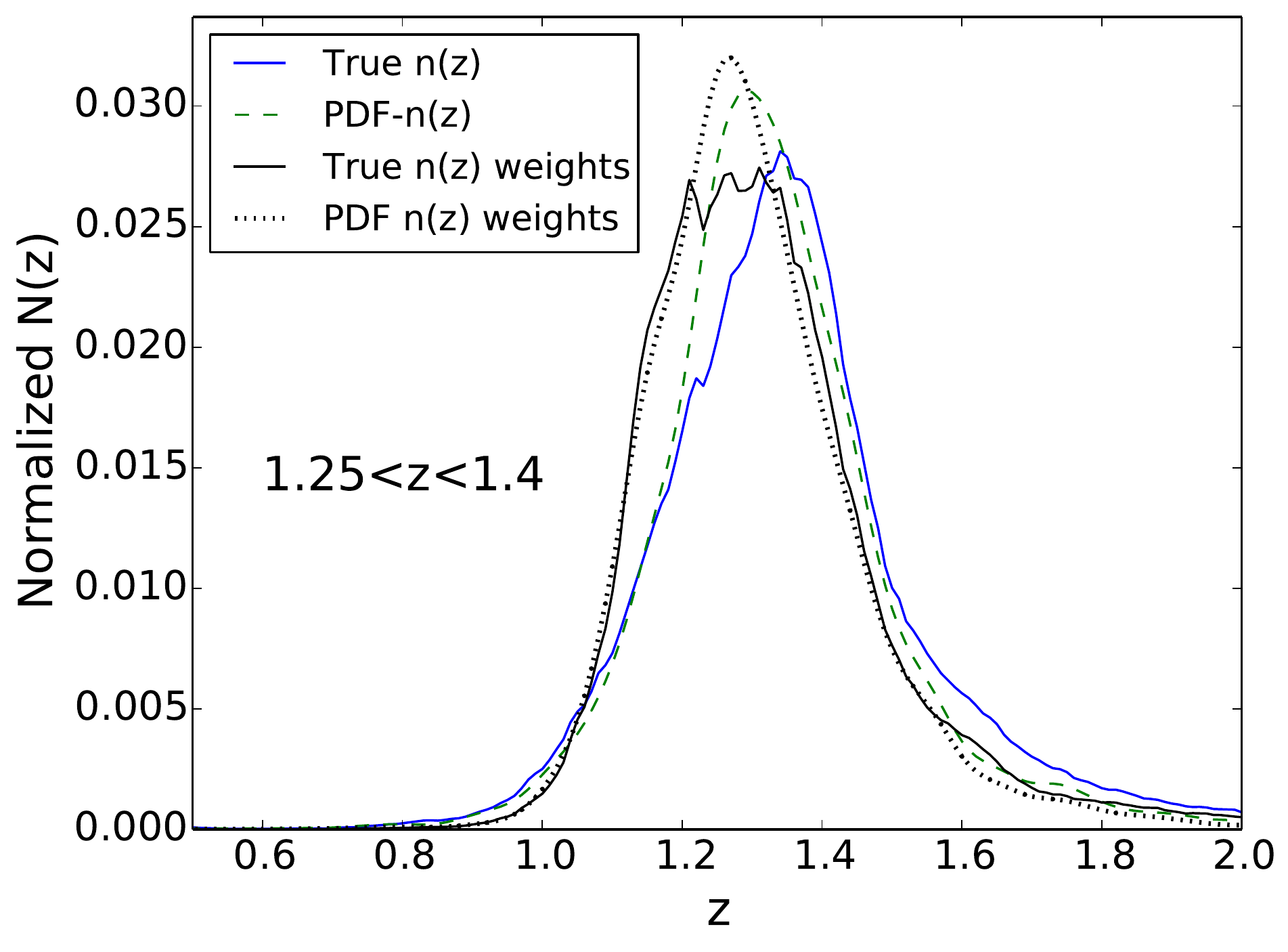}
\includegraphics[trim = 0cm 0cm 0cm 0cm, width=0.49\textwidth]{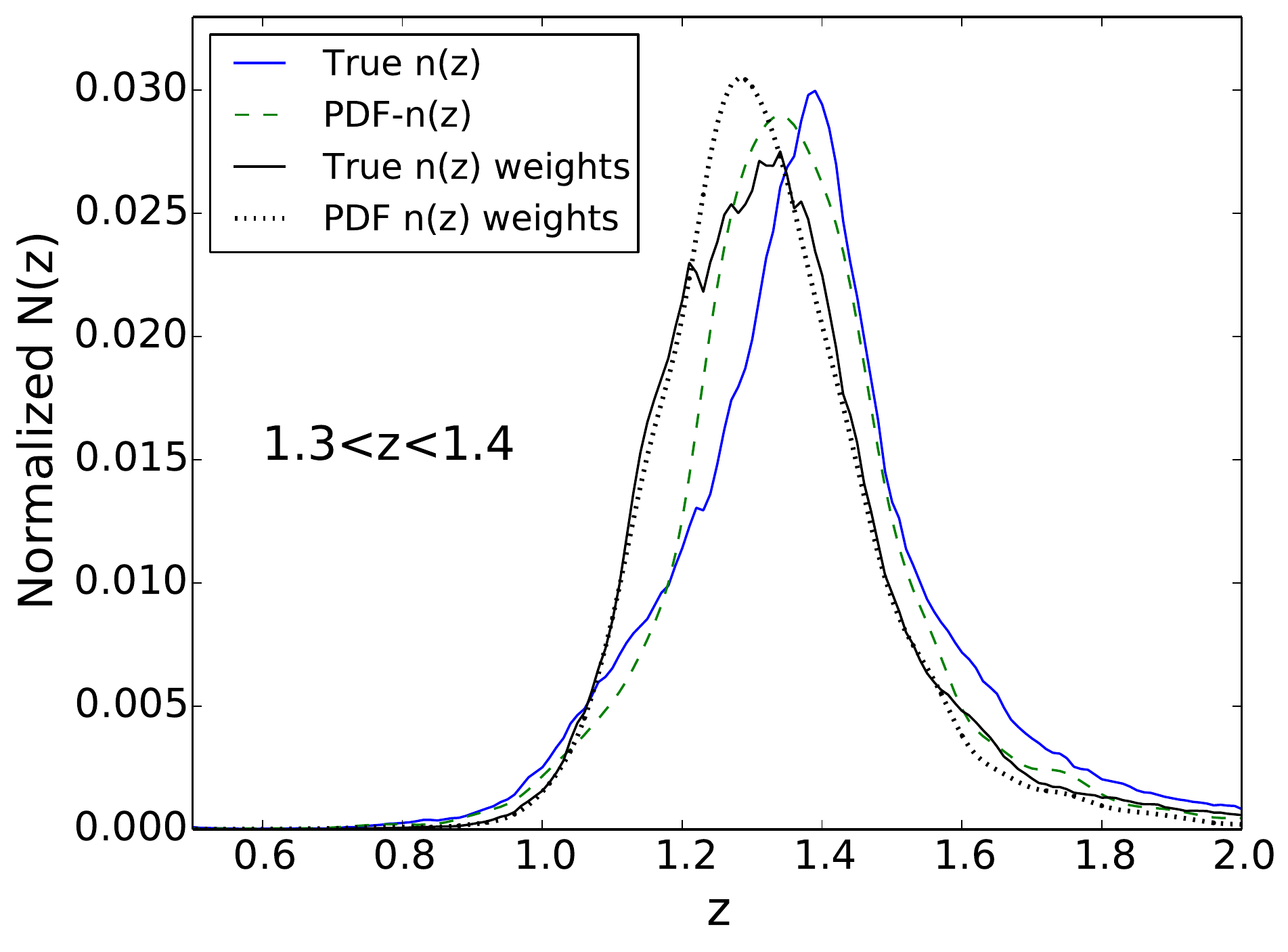}
\caption{\emph{High redshift}: A comparison between the true redshift distribution of galaxies selected according to mean photo-z redshift compared with the distribution given by the PDF stacking of the same sample over the highest redshift range of the true galaxy sample. We also show the redshift distributions when stacking weighted PDFs (solid black) and true redshifts of weighted galaxies (dotted black) for the different redshift bins.} 
 
\
\label{fig:figap5}
\end{center}
\end{figure*}

the true redshift distribution. This disagreement is expected and has been observed when using template based and machine learning algorithms that incorporate PDFs. We expanded the comparison to intermediate redshift (Figure \ref{fig:figap4}) and high redshift (Figure \ref{fig:figap5}), observing similar differences. For comparison, we also show the redshift distribution for the PDF weighted galaxies, when stacking PDFs or true redshifts.

\bsp

\label{lastpage}

\end{document}